\pdfoutput=1
\pdfmapfile{+mtpro2.map}
\documentclass[%
reprint,
twocolumn,
aps, prb,
superscriptaddress,
]{revtex4-2}

\usepackage{newtxtext}

\IfFileExists{mtpro2.sty}{%
\usepackage[subscriptcorrection]{mtpro2}%
\pdfmapfile{+ ./texmf/fonts/map/dvips/mtpro2/mtpro2.map}%
}{%
\usepackage{newtxmath}%
}

\usepackage{graphicx}
\usepackage{placeins}

\usepackage{physics}
\usepackage[version=3]{mhchem}

\usepackage[dvipsnames]{xcolor}
\definecolor{ColumbiaBlue}{RGB}{23, 66, 150}

\usepackage[T1]{fontenc}
\usepackage{hyperref}
\usepackage{etoolbox}
\hypersetup{
    colorlinks,
    linkcolor={Maroon},
    citecolor={ForestGreen},
    urlcolor={ColumbiaBlue}
}

\makeatletter
\patchcmd{\NAT@citexnum}{--\NAT@penalty}{-\NAT@penalty}{}{}
\makeatother

\newcommand{\ZHC}[1]{#1}

\newcommand{\REV}[1]{#1}

\newcommand*{\ii}{{\mathrm{i}}}
\newcommand*{\ee}{{\mathrm{e}}}
\newcommand*{\diff}{ {\mathrm{d}} }

\newcommand{\veck}{{\mathbf{k}}}
\newcommand{\vecq}{{\mathbf{q}}}

\newcommand{\vecp}{{\mathbf{p}}}

\newcommand{\etal}{\textit{et al.}}

\newcommand*{\abinitio}{{\textit{ab initio}} }

\newcommand{\uroman}[1]{\uppercase\expandafter{\romannumeral#1}}
\newcommand{\lroman}[1]{\romannumeral #1}

\begin{document}

\title{Self-consistent GW theory for superconductivity in \texorpdfstring{\ce{SrTiO3}}{SrTiO3} models}

\author{Zhi-Hao Cui}
 \email{zhcui0408@gmail.com}
\affiliation{Department of Chemistry, Columbia University, New York, New York 10027, USA}
\affiliation{Department of Chemistry, University of California, Irvine, CA 92697, USA}

\author{John Sous}
\email{john.sous@yale.edu}
\affiliation{Department of Applied Physics, Yale University, New Haven, Connecticut 06511, USA}
\affiliation{Energy Sciences Institute, Yale University, West Haven, Connecticut 06516, USA}

\author{Andrew J. Millis}
 \email{ajm2010@columbia.edu}
\affiliation{Department of Physics, Columbia University, New York, New York 10027, USA}
\affiliation{Center for Computational Quantum Physics, Flatiron Institute, 162 5th Avenue, New York, New York 10010, USA}

\author{David R. Reichman}
 \email{drr2103@columbia.edu}
\affiliation{Department of Chemistry, Columbia University, New York, New York 10027, USA}

\pdfbookmark[0]{Main Text}{Main Text}

\begin{abstract}
Superconductivity in doped $\mathrm{SrTiO}_3$ occurs \REV{over a wide range of} carrier densities \REV{including those} for which the Fermi energy is below the polar longitudinal optical phonon scale. In this regime, \REV{the assumptions underpinning the conventional} implementations of Migdal-Eliashberg theory, including frequency cutoffs \REV{at the phonon scale}  and a Coulomb pseudopotential $\mu^\ast$, are not \REV{valid}. We solve \REV{the finite-temperature} $GW$ equations \REV{including the full momentum and frequency dependence without cutoffs or $\mu^\ast$} for polar one-band models of $\mathrm{SrTiO}_3$, with effective masses and three-phonon dielectric functions parameterized from \abinitio calculations.  Comparing different self-consistency levels, namely $G_0W_0$, $GW_0$, and fully self-consistent $GW$, we find that the one-shot \REV{($G_0W_0$)} kernel overestimates the pairing-onset \REV{temperature} by one to two orders of magnitude. The dominant suppression comes from \REV{changing $G_0$ to $G$, thereby incorporating the phonon renormalization factor in the electron Green function.} \REV{Using the self-consistently computed interaction} $W$ further lowers and narrows the pairing-onset dome. \REV{In the dilute limit,} our calculations identify the  \REV{pairing channel as the Fr\"ohlich phonon interaction screened by the incipient ferroelectricity of the material, with plasmonic and electronic screening effects negligible.}  The \REV{numerical solution of the full equations reveals a pairing-onset scale that remains non-zero as the density tends to zero}, whereas Fermi-surface projection or Fermi-energy frequency truncation removes it. This work highlights the relevance of \REV{incipient ferroelectricity, the importance of} self-consistency \REV{and the need for a full momentum and frequency dependent treatment}
in the modeling of superconductivity in \ce{SrTiO3}-like doped polar semiconductors.

\end{abstract}

\maketitle

\section{Introduction}\label{sec:intro}

Doped \ce{SrTiO3} (STO) was the first superconducting semiconductor~\cite{Schooley64}, and sixty years later its superconductivity remains incompletely understood.  Its  \REV{critical temperature traces a dome with a maximum $T_{\mathrm{c}} \approx 0.4$~K near $n \approx 10^{20}\,\mathrm{cm}^{-3}$~\cite{Koonce67,Lin14,Collignon19}, the} transition survives down to carrier densities of order $10^{17}\mathrm{cm}^{-3}$, \REV{i.e. $\sim 10^{-5}$ per unit cell}~\cite{Lin13}. The  microwave and tunneling experiments have sharpened the puzzle by showing single-gap electrodynamics and BCS-like gap ratios \REV{even at carrier densities such that the Fermi energy is small compared to the relevant phonon frequencies}~\cite{Thiemann18,Yoon21}.  Isotope and calcium-tuned experiments further connect the dome to the vicinity of ferroelectric order~\cite{Rischau17-ferroelectric,Rischau22-isotope}.  

The material \REV{exhibits} features that \REV{call into question the applicability of} the standard theory of phonon-mediated pairing~\cite{Gastiasoro20}.  First, the dominant electron-phonon coupling is the long-range Fr\"ohlich interaction with the polar longitudinal optical (LO) modes, singular as $1/q$ at small momentum transfer.  Second, over most of the dome the Fermi energy is far below the \REV{frequency of the LO phonons}, and thus the phonons are antiadiabatic, and spectroscopic probes \REV{indicate that the electrons are strongly renormalized by electron-phonon coupling}~\cite{VanMechelen08,Devreese10-STO-polaron,Swartz18}.  Third, the Coulomb repulsion cannot be folded into a Morel-Anderson pseudopotential $\mu^\ast$~\cite{MorelAnderson62,Bogoliubov59}, because the logarithmic reduction arising from $\ln(E_{\mathrm{F}}/\omega_{\mathrm{LO}})$ requires $E_{\mathrm{F}} \gg \omega_{\mathrm{LO}}$; in dilute STO the inequality runs the other way.

These difficulties have produced many proposed mechanisms but comparatively few controlled numerical calculations. Pairing has been attributed to the LO phonons themselves, including dielectric-function and multiband implementations~\cite{Koonce67,Gorkov16,Klimin16-STO-LO,Klimin19-STO-dielectric}, to the plasmon or, more generally, the dynamically screened Coulomb interaction~\cite{Takada78-STO,Takada80-STO,RuhmanLee16,Enderlein20-polar-modes}, and to the soft transverse optical (TO) mode associated with the incipient ferroelectric transition~\cite{Appel69,Edge15,Wolfle18-STO,Marel19,Volkov22,Yu22-quantum-paraelectrics,Saha25-strong-coupling}.  Within any one of these proposals, quantitative answers depend on technical choices (for example, Fermi-surface projection versus the full momentum dependence, frequency cutoffs, pseudopotentials, etc.) whose consequences in the antiadiabatic regime \REV{require further examination}.  Analytic work on the dilute limit has shown that these choices are not innocuous: the full momentum gap equation has a finite pairing-onset scale as the carrier density $n \to 0$~\cite{Gastiasoro19-lowdensity,PhanChubukov22-lowdensity}, whereas Fermi-surface estimates of the same model give an exponentially small scale.

In this paper we address these questions for polar one-band models of doped STO, in which a parabolic band interacts through a single dynamical interaction $\mathcal{V}(q,\ii\nu_n)$ that combines the Coulomb repulsion and the three polar-phonon branches.  We solve the finite-temperature $GW$ equations~\cite{Hedin65,Hybertsen86,Aryasetiawan98RPP,Golze19-GWCompendium,Li20-sparse-grids-gw-gf,Yeh22-finite-temp-scGW}; for the bare-vertex interaction considered here, these equations are the self-consistent, phonon-renormalized Migdal-Eliashberg approximation~\cite{Marsiglio90-migdal-renorm,Marsiglio20} written in terms of the screened interaction $W$.  We then solve the associated linearized pairing eigenvalue problem, keeping the full momentum and Matsubara-frequency structure throughout.  Related one-loop treatments have emphasized the same full momentum- and frequency-dependent structure in coupled phonon-plasmon pairing problems~\cite{Veld23-screening}.  The calculations are numerically converged in all momentum and frequency variables, which makes them a controlled reference for the vertex-free theory of this model, against which further approximations can be tested one at a time.  Specifically, we answer four questions:

(\lroman{1})~\emph{How much does self-consistency matter?}  We compare one-shot $G_0W_0$, partially self-consistent $GW_0$ (dressed $G$, fixed screening), and fully self-consistent sc$GW$ (dressed $G$ and $W$).  We find that normal-state self-consistency strongly suppresses the one-shot pairing kernel, mainly through the Eliashberg renormalization $Z$; updating $W$ further lowers and narrows the onset dome.

(\lroman{2})~\emph{What mediates the pairing?}  Decomposition calculations in which we switch off either the electronic polarization or the phonon dielectric factor identify the \REV{Fr\"ohlich electron-phonon coupling, modified by the screening associated with incipient ferroelectricity} as the low-density pairing channel.  

(\lroman{3})~\emph{What happens in the high- and low-density limits?}  At high density the Fermi-surface-averaged kernel weakens as $\sim \ln k_{\mathrm{F}}/k_{\mathrm{F}}$ and the onset collapses exponentially; at low density the \REV{pairing-onset temperature predicted by the} full gap equation approaches a \REV{non-zero}, weakly density-dependent value, consistent with the analytic results of Refs.~\cite{Gastiasoro19-lowdensity,PhanChubukov22-lowdensity}.

(\lroman{4})~\emph{Which technical shortcuts fail, and where?}  The full calculation retains a \REV{non-zero} dilute-limit pairing onset because, when $E_{\mathrm{F}}$ lies below the polar LO pole window, the pairing weight lives through the LO frequency range and extends to momenta $k_{\mathrm{LO},j} = \sqrt{2m^\ast \omega_{\mathrm{LO},j}} \gg k_{\mathrm{F}}$.  Truncating the Matsubara window of the pairing kernel at the Fermi energy, or projecting the kernel onto the Fermi surface in the manner of Takada~\cite{Takada80-STO}, discards exactly this weight and eliminates the dilute-limit pairing entirely.

The remainder of the paper is organized as follows: Section~\ref{sec:model method} defines the model and the hierarchy of $GW$ approximations and summarizes the numerical approach; implementation details are collected in the Appendices.  Section~\ref{sec:results} presents the results \REV{in terms of} the four questions above.  Section~\ref{sec:discussion} discusses the physics left out of the model, and Sec.~\ref{sec:conclusion} concludes. \REV{Appendices present technical specifics and consistency checks}.

\vspace{-0.2cm}
\section{Model and methods}\label{sec:model method}

\vspace{-0.2cm}
\subsection{Polar one-band models}\label{subsec:model-def}

We model the conduction band of doped STO by a single isotropic parabolic band, populated by a carrier density $n$.  In atomic units,
\begin{equation}
\varepsilon_{\veck} = \frac{k^2}{2m^\ast},\qquad
k_{\mathrm{F}} = (3\pi^2 n)^{1/3},
\label{eq:dispersion}
\end{equation}
with $m^\ast = 2.1\,m_{\mathrm{e}}$ the density-of-states mass appropriate to the $t_{2g}$ conduction-band bottom (Appendix~\ref{app:abinitio}).  The carriers interact \REV{through the Coulomb interaction and couple to polar lattice vibrations through the usual Fr\"ohlich interaction. Density functional perturbation theory (DFPT) calculations confirm that at the densities of interest, coupling to other phonon modes is negligible.}  We absorb the \REV{three} zone-center polar branches into the lattice dielectric factor on the imaginary axis, \REV{so that the} interaction used in the main calculations is
\begin{equation}
\mathcal{V}(q, \ii\nu_n)
=\frac{4\pi}{\epsilon_\infty q^2}\,F_3(\nu_n),
\qquad
F_3(\nu)=
\prod_{j=1}^{3}
\frac{\nu^2+\omega_{\mathrm{TO},j}^{2}}
     {\nu^2+\omega_{\mathrm{LO},j}^{2}},
\label{eq:Vtot}
\end{equation}
where $\ii\nu_n = 2n\pi T$ are bosonic Matsubara frequencies \REV{and $\epsilon_\infty$ is the dielectric function at frequencies above the highest phonon frequency and below the first electronic interband transition}.  The generalized Lyddane-Sachs-Teller relation~\cite{Lyddane41,ChavesPorto73-gLST} \REV{relates the static limit of $F_3$ to the zero-frequency dielectric enhancement $\epsilon_0$:}
\begin{equation}
F_3(0)=\frac{\epsilon_\infty}{\epsilon_0},
\qquad
\frac{\epsilon_0}{\epsilon_\infty}
=\prod_{j=1}^{3}
\frac{\omega_{\mathrm{LO},j}^{2}}
     {\omega_{\mathrm{TO},j}^{2}},
\end{equation}

\REV{In a nearly ferroelectric system the large contrast between $\epsilon_0$ and $\epsilon_\infty$ means that the electron-lattice interaction is almost vanishing at low frequencies}  and returns to the $\epsilon_\infty$-screened Coulomb interaction at high frequency.
 \REV{The change in magnitude as the frequency is varied through the LO frequencies implies an} attraction-like retardation structure even though the interaction is repulsive for every Matsubara frequency.

The numerical parameters \REV{used in this work} are $\epsilon_\infty = 6.16$, $\epsilon_0 = 2.3\times 10^4$, and the three-mode pole set
\begin{equation}
\begin{aligned}
\omega_{\mathrm{TO}} &= (1.24,\;22.39,\;73.24)~\mathrm{meV},\\
\omega_{\mathrm{LO}} &= (21.80,\;55.72,\;102.14)~\mathrm{meV}.
\end{aligned}
\end{equation}
The lowest TO pole is anchored by the experimental low-temperature static dielectric constant ($\epsilon_0$) of quantum-paraelectric STO~\cite{MullerBurkard79}, while the higher TO poles and all LO zeros are extracted from the DFPT/temperature-dependent effective potential construction (TDEP) calculations described in Appendix~\ref{app:abinitio}. \REV{As written, the theory neglects the dependence of $\omega_{\mathrm{TO}}$ on density and wave vector; Appendix~\ref{app:fauque-softmode} } \REV{shows that use of the experimentally measured ~\cite{Fauque25-dipolar-length} density and momentum dependence of the TO phonon frequency does not change the results significantly.}

\REV{We may define an approximately equivalent effective one-mode model by rewriting
\begin{equation}
F_3(\ii\nu)=1-\sum_j C_j\frac{\omega_{\mathrm{LO},j}^2}{\nu^2+\omega_{\mathrm{LO},j}^2} ,
\label{eq:Cjdef}
\end{equation}
where the $C_j$ are determined by $\omega_{\mathrm{TO},j}/\omega_{\mathrm{LO},j}$. We then define an effective LO phonon frequency via
\begin{equation}    \ln\Omega_{\mathrm{eff}}=\frac{\sum_j C_j \ln\omega_{\mathrm{LO},j}}{\sum_j C_j} ,
    \label{eq:Omegaeff}
    \end{equation}
and an effective TO frequency  $\omega_{\mathrm{TO}, \mathrm{eff}}=\Omega_{\mathrm{eff}}\sqrt{\epsilon_\infty/\epsilon_0}$.  In the one-pole limit Eq.~\eqref{eq:Vtot} reduces to the dielectric form used in earlier pairing theories~\cite{BardeenPines55,Kirzhnits73-dielectric,Takada80-STO}.}
\REV{The three pole parameters imply} the log-strength compressed scale $\Omega_{\mathrm{eff}}=84.26$~meV and the LST-consistent $\omega_{\mathrm{TO,eff}}=\Omega_{\mathrm{eff}}\sqrt{\epsilon_\infty/\epsilon_0}=1.38$~meV.  When discussing analytic limits we use a one-scale notation $\omega_0$ only as an asymptotic shorthand for such a polar crossover scale; the full numerical results always use the product form above.

\subsection{\texorpdfstring{$GW$}{GW} equations and the linearized gap problem}\label{subsec:gw-def}

We work on the Matsubara axis at temperature $T$, with fermionic frequencies $\ii\omega_m = (2m+1)\pi T$.  The normal-state Green's function is
\begin{equation}
G(\veck, \ii\omega_m) = \frac{1}{\ii\omega_m - \qty[\varepsilon_{\veck} - \mu + \Sigma(\veck, \ii\omega_m)]},
\label{eq:elec GF}
\end{equation}
\REV{and the bare Green function $G_0$ is $G$ with the self-energy $\Sigma$ removed.
We take the interaction $W$ to be the bare interaction $\mathcal{V}$ (Eq.~\eqref{eq:Vtot}) screened by  the electronic polarization operator $\Pi$:
\begin{equation}
W(\vecq, \ii \nu_n) = \frac{\mathcal{V}(\vecq, \ii \nu_n)}{1 - \mathcal{V}(\vecq, \ii \nu_n)\, \Pi(\vecq, \ii \nu_n)},
\label{eq:effective W}
\end{equation}
and for $\Pi$ we take the random-phase approximation (RPA) form in which} the particle-hole polarization is \REV{approximated by} the $GG$ bubble,
\begin{equation}
\Pi(\vecq, \ii\nu_n) = \frac{2}{\beta} \sum_{m} \int_{\veck}  G(\veck, \ii \omega_m)\, G(\veck - \vecq, \ii \omega_m - \ii \nu_n).
\label{eq:Pi}
\end{equation}

With this Matsubara convention the static Lindhard function is negative, so in the static metallic limit the denominator becomes $1+\mathcal{V}|\Pi(q,0)|$.

The self-energy is the $GW$ exchange diagram,
\begin{equation}
\Sigma(\veck, \ii \omega_m) = -\frac{1}{\beta} \sum_{n} \int_{\vecq}  G(\veck - \vecq, \ii \omega_m - \ii \nu_n)\, W(\vecq, \ii \nu_n),
\label{eq:effective GW}
\end{equation}
with $\int_{\vecq} \equiv \int \diff^3 q/(2\pi)^3$ and $\beta = 1/T$.  

\REV{Eqs.~\eqref{eq:effective W}-\eqref{eq:effective GW} are solved iteratively from an initial guess for $G$. At each step in the iteration} the chemical potential $\mu$ is adjusted so that $G$ reproduces the target density $n$.  \REV{This set of self-consistent equations is equivalent to the \emph{renormalized} Migdal-Eliashberg theory of Marsiglio~\cite{Marsiglio90-migdal-renorm}.}  The explicit connection is given in Appendix~\ref{app:equivalence}.

The strength of the \REV{interaction may be parametrized} by the Eliashberg mass-renormalization function
\begin{equation}
	Z(\veck, \ii\omega_m) = 1 - \frac{\Sigma(\veck,\ii\omega_m) -
	\Sigma(\veck,-\ii\omega_m)}{2\,\ii\omega_m},
\label{eq:Z}
\end{equation}
which we quote at $k = k_{\mathrm{F}}$ and the lowest fermionic frequency, $Z(k_{\mathrm{F}}, \ii\pi T)$.  This is the Matsubara-axis $Z$ of Eliashberg theory, not the real-axis quasiparticle residue $Z_{\mathrm{qp}}$; on the Fermi surface and at low frequency the two are reciprocal, $Z_{\mathrm{qp}} \approx 1/Z$.

Superconductivity is diagnosed from the linearized anomalous self-energy.  Writing the anomalous vertex as $\Phi(\veck,\ii\omega_m)$ and linearizing in $\Phi$ gives the eigenvalue problem
\begin{equation}
\begin{split}
\lambda(T)\,\Phi(\veck', \ii\omega_{m'})
= -\frac{1}{\beta} \sum_{m}\!\int_{\veck} &W(\veck-\veck', \ii\omega_m - \ii\omega_{m'})\, \times \\
& \abs{G(\veck, \ii\omega_m)}^2\, 
\Phi(\veck, \ii\omega_m),
\end{split}
\label{eq:LGE}
\end{equation}
where $\abs{G}^2 \equiv G(\veck,\ii\omega_m)\,G(-\veck,-\ii\omega_m)$ for the isotropic band considered here.  The leading eigenvalue $\lambda(T)$ grows on cooling, and the temperature at which $\lambda(T_{\mathrm{c}}) = 1$ defines the \emph{pairing-onset} temperature.  We emphasize the name: Eq.~\eqref{eq:LGE} signals the instability of the normal state toward pair formation, whereas long-range superconducting order also requires phase stiffness.  Only in the degenerate regime, where the stiffness scale is comparable to or larger than the pairing scale, can this crossing be identified with the superconducting transition; in the dilute nondegenerate limit the physical transition is bounded instead by \REV{ $T_\theta \sim E^*_{\mathrm{F}}\equiv k_{\mathrm{F}}^2/(2 Z m^\ast)$} (Sec.~\ref{subsec:tc-limits}).  We solve Eq.~\eqref{eq:LGE} over the full radial momentum grid and the full sparse-IR Matsubara grid: no Fermi-surface projection, no frequency window, and no Coulomb pseudopotential $\mu^\ast$ enter unless specified.

\begin{table}[!htb]
\vspace{-0.2cm}
\caption{Hierarchy of approximations compared in this work.  All methods solve the same linearized gap equation~\eqref{eq:LGE}; they differ in the propagator and screening that build the kernel.  $\Pi_0$ denotes the Lindhard function of the bare band.  These methods are used in Sec.~\ref{subsec:mechanism} to decompose the pairing mechanism.
\label{tab:methods}}
\begin{ruledtabular}
\begin{tabular}{llll}
Label & $G$ & $\Pi$ in $W$ & Role \\
\hline
$G_0W_0$ & bare $G_0$ & $\Pi_0$, fixed & one-shot \\
$GW_0$ & self-consistent & $\Pi_0$, fixed & dressed $G$ \\
sc$GW$ & self-consistent & $GG$, updated & full feedback \\
\hline
$G_0\mathcal{V}$ & bare $G_0$ & $\Pi \equiv 0$ & lattice only \\
$G\mathcal{V}$ & self-consistent & $\Pi \equiv 0$ & dressed, lattice only \\
no phonon & as labeled & as labeled & pure Coulomb \\
\end{tabular}
\end{ruledtabular}
\end{table}

\REV{As outlined in Table~\ref{tab:methods}, we consider a hierarchy of approximations to the full set of Eqs.~\eqref{eq:effective W}- \eqref{eq:effective GW}~\cite{Stan09}.}  Related quasiparticle self-consistent schemes instead iterate an effective quasiparticle Hamiltonian~\cite{Faleev04,Lei22-Gaussian-QSGW}.  $G_0W_0$ evaluates Eqs.~\eqref{eq:effective W}-\eqref{eq:effective GW} once, with the bare propagator and the Lindhard polarization $\Pi_0$ \REV{computed from the bare Green function $G_0$}~\cite{Hybertsen85,Jiang16a}.  $GW_0$ iterates the Dyson equation~\eqref{eq:elec GF} and the density constraint to self-consistency in $G$ at fixed $W_0$.  sc$GW$ additionally rebuilds $\Pi$ from the dressed $G$ at every iteration~\cite{Holm98,Caruso12,Grumet18-full-scGW}, so that the screening and the phonon renormalization feed back on the pairing kernel.
The $G_0\mathcal{V}$ and $G\mathcal{V}$ decompositions remove the electronic (plasmon) screening while keeping the lattice dielectric factor, at the one-shot and dressed-propagator levels respectively;
the no-phonon variant sets $F_3(\nu)=1$, which makes $\mathcal{V}$ a pure $\epsilon_\infty$-screened Coulomb interaction whose only frequency dependence is generated by $\Pi$.

\vspace{-0.2cm}
\subsection{Numerical approach}\label{subsec:numerics}

The isotropy of the model allows every momentum convolution to be reduced to coupled 1D radial integrals, with the Coulomb $1/q^2$ head of the kernel integrated analytically rather than sampled.  Frequency sums are represented with the sparse intermediate-representation (IR) basis~\cite{Shinaoka17,Li20-sparse-grids-gw-gf,Wang23-EliashbergIR}; in production we use an IR cutoff $\Lambda=10^9$, retaining the high-frequency tails of $W$ and $G$ explicitly and scanning the temperature down to $0.1$~K.  The gap equation~\eqref{eq:LGE} is solved as a Davidson eigenproblem on the same grids.  Implementation details and convergence tests are given in Appendices~\ref{app:implementation} and~\ref{app:conv}; the open-source implementation is available in \textsc{migdal}~\cite{MigdalCode}.

\begin{figure*}[!htb]
\includegraphics[width=0.98\textwidth,clip]{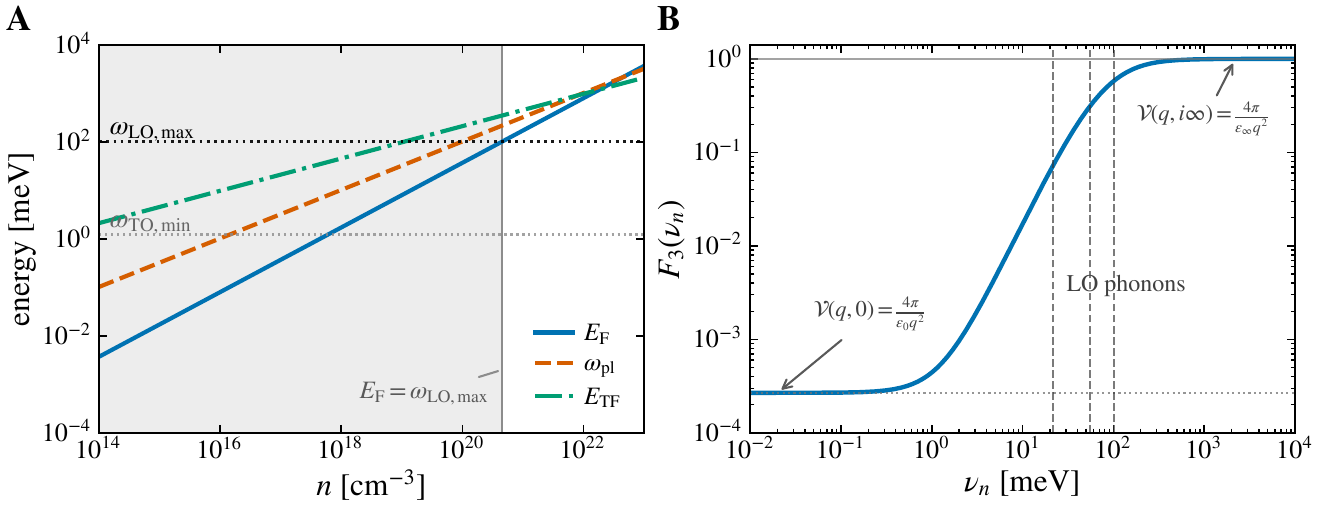}
\caption{(A)~Electronic energy scales of the three-mode model across the density range, for $m^\ast = 2.1 m_{\mathrm{e}}$ and $\epsilon_\infty = 6.16$: Fermi energy $E_{\mathrm{F}}$, plasma frequency $\omega_{\mathrm{pl}} = (4\pi n/\epsilon_\infty m^\ast)^{1/2}$, and Thomas-Fermi energy $E_{\mathrm{TF}} = q_{\mathrm{TF}}^2/2m^\ast$, compared with the softest TO and highest LO reference scales of Eq.~\eqref{eq:Vtot}.  The vertical line marks $E_{\mathrm{F}} = \omega_{\mathrm{LO,max}} = 102.14$~meV at $n^\ast \approx 4.51\times10^{20}\,\mathrm{cm}^{-3}$.  (B)~Lattice dielectric factor $F_3(\nu_n)$ multiplying the $\epsilon_\infty$-screened Coulomb interaction in Eq.~\eqref{eq:Vtot}.  Arrows mark the static and high-frequency limits, $\mathcal{V}(q,0)=\frac{4\pi}{\epsilon_0 q^2}$ and $\mathcal{V}(q,\ii\infty)=\frac{4\pi}{\epsilon_\infty q^2}$; dashed vertical lines mark the LO poles at $21.80$, $55.72$, and $102.14$~meV.
\label{fig:scales}}
\end{figure*}
\section{Results}\label{sec:results}

\subsection{Energy scales and regimes}\label{subsec:scales}

Fig.~\ref{fig:scales} organizes the discussion.  Panel~(A) compares the density-dependent electronic scales with the fixed polar-phonon scale.  The Fermi energy crosses the LO pole range (21.80 - 102.14~meV) \REV{as the density is increased through the range }$4.45\times10^{19}$ and $4.51\times10^{20}\,\mathrm{cm}^{-3}$, so the entire low-density flank of the experimental dome is antiadiabatic with respect to most of the polar spectral weight.  The plasma frequency crosses the same LO range between $4.46\times10^{18}$ and $9.79\times10^{19}\,\mathrm{cm}^{-3}$, so the importance of electronic screening changes across the dome.  Panel~(B) shows the corresponding frequency profile of the bare interaction factor: $F_3(0)=\epsilon_\infty/\epsilon_0=2.68\times10^{-4}$ at the static end, while $F_3(\nu)\to1$ at high frequency, with the main crossover occurring through the LO pole window.  This is the scale over which the kernel changes from the statically lattice-screened limit, $\mathcal{V}(q,0)=\frac{4\pi}{\epsilon_0 q^2}$, to the Coulomb-dominated limit, $\mathcal{V}(q,\ii\infty)=\frac{4\pi}{\epsilon_\infty q^2}$.   \REV{Two} consequences follow immediately. \REV{At densities less than $\sim 10^{20}$, the basic interaction $\mathcal{V}$ is very small at the scale of the Fermi energy; in this regime any approximation that restricts attention only to the vicinity of the Fermi energy will miss the important interaction effects. Conversely, only at densities $\gtrsim 10^{22}$ is the Fermi energy much larger than the phonon frequency; so only at this high density range is the standard Tolmachev-Morel-Anderson theory of the Coulomb pseudopotential applicable. } 

\subsection{Effect of self-consistency}\label{subsec:selfconsistency}

\begin{figure*}[!htb]
\includegraphics[width=0.98\textwidth,clip]{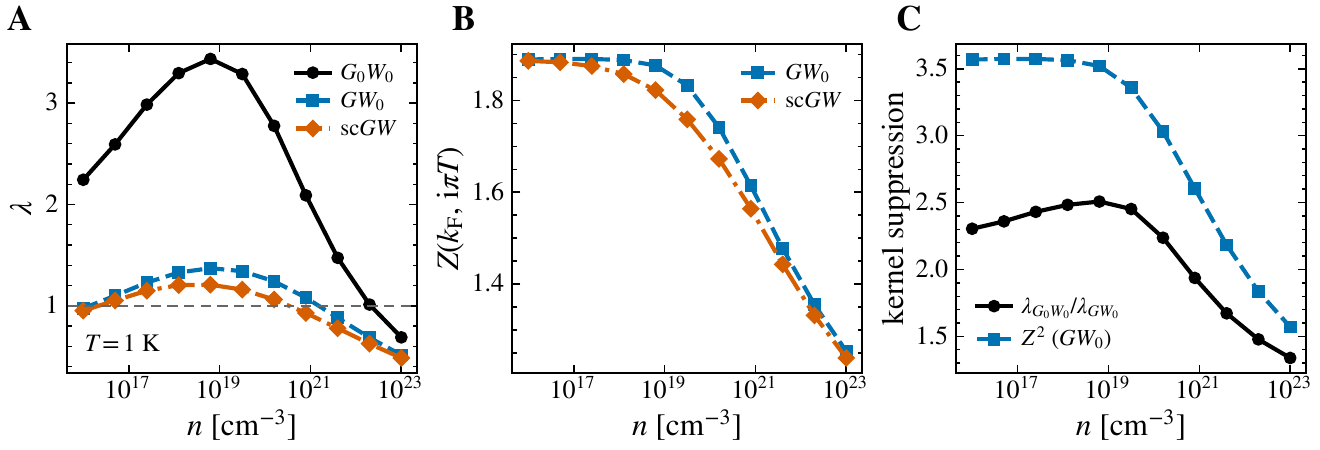}
\caption{(A)~Leading eigenvalue $\lambda$ of the linearized gap equation at $T = 1$~K as a function of carrier density for the one-shot ($G_0W_0$), partially self-consistent ($GW_0$), and fully self-consistent (sc$GW$) kernels.  The dashed line marks the instability condition $\lambda = 1$.  (B)~Eliashberg renormalization function $Z(k_{\mathrm{F}}, \ii\pi T)$ of Eq.~\eqref{eq:Z} for the two self-consistent methods.  (C)~Kernel-suppression analysis: the eigenvalue ratio $\lambda_{G_0W_0}/\lambda_{GW_0}$ compared with $Z^2$ of the $GW_0$ solution.  $Z^2$ gives the expected near-Fermi-surface suppression scale and reproduces the density dependence.
\label{fig:hierarchy}}
\end{figure*}

Fig.~\ref{fig:hierarchy}(A) shows the leading pairing eigenvalue at $T = 1$~K across seven decades of density.  The one-shot $G_0W_0$ kernel gives a broad instability, with $\lambda$ rising from $2.24$ at $10^{16}\,\mathrm{cm}^{-3}$ to a maximum $3.44$ near $6.3\times10^{18}\,\mathrm{cm}^{-3}$ before falling on the high-density side.  Dressing the propagator at fixed screening ($GW_0$) collapses the antiadiabatic-regime eigenvalues to roughly $1.0$-$1.37$, a reduction by a factor of about $2$-$2.6$.  Updating the screening as well (sc$GW$) gives a further, density-dependent suppression, largest near and above the dome maximum where electronic polarization is most active.  Because the onset temperature depends exponentially on the kernel strength, this eigenvalue shift is far from negligible: it lowers the peak onset and suppresses the high-density flank by substantially larger factors (Sec.~\ref{subsec:tc-limits}).

The mechanism of the reduction can be read off from Figs.~\ref{fig:hierarchy}(B) and (C).  \REV{As density is decreased,} the Eliashberg renormalization $Z(k_{\mathrm{F}}, \ii\pi T)$ grows from $1.25$ at $n = 10^{23}\,\mathrm{cm}^{-3}$ \REV{and saturates at about}  $1.9$ in the dilute limit. The corresponding quasiparticle weight is $Z_{\mathrm{qp}} \approx 1/Z \approx 0.5$. \REV{Thus the dilute limit may be described as a Fermi liquid of quasiparticles, substantially dressed by phonons, with the dressing coming from high frequencies (relative to the Fermi energy).}  The pairing kernel of Eq.~\eqref{eq:LGE} contains $\abs{G}^2$, which on the Fermi surface and at the lowest frequency is suppressed by exactly $1/Z^2$.  Panel~(C) tests how far this single factor carries: the eigenvalue ratio $\lambda_{G_0W_0}/\lambda_{GW_0}$ reaches $2.3$-$2.6$ in the antiadiabatic regime and follows the density dependence of $Z^2 \approx 3.6$, which lies above it.  The simple estimate is not saturated because only the low-frequency, near-Fermi-surface part of the kernel is suppressed by the full $1/Z^2$; the high-frequency and off-Fermi-surface weight, which carries a substantial part of the dilute pairing (Sec.~\ref{subsec:projection}), is renormalized more weakly, and the self-consistent chemical-potential shift and the momentum structure of $\Sigma$ contribute at the same order.  The suppression \REV{of $T_{\mathrm{c}}$} is not an artifact of the pairing calculation: it is the statement that the carriers that pair are  \REV{strongly dressed by phonons}.  Complementary normal-state analyses of the doped Fr\"ohlich solid using Fan--Migdal Dyson and second-order cumulant approaches show how finite band filling and Lindhard free-carrier screening reshape the density-dependent mass renormalization, while also identifying parameter regimes in which these low-order treatments cease to yield a well-defined quasiparticle picture~\cite{Kandolf22-doped-Frohlich}.

Two further observations.  First, the screening update acts most strongly where the dome closes: the high-density edge of the $\lambda(1\,\mathrm{K}) > 1$ window moves in from $n \approx 2\times10^{22}\,\mathrm{cm}^{-3}$ at the $G_0W_0$ level to $\approx 8\times10^{20}\,\mathrm{cm}^{-3}$ for $GW_0$ and $\approx 1.6\times10^{20}\,\mathrm{cm}^{-3}$ for sc$GW$, so each stage of self-consistency narrows the computed dome further.  Second, at the lowest densities $GW_0$ and sc$GW$ nearly coincide, \REV{because in this regime} the electronic polarization is negligible and the kernel reduces to the lattice-screened interaction of Sec.~\ref{subsec:mechanism}.

\subsection{What mediates the pairing}\label{subsec:mechanism}

\begin{figure*}[!htb]
\includegraphics[width=0.98\textwidth,clip]{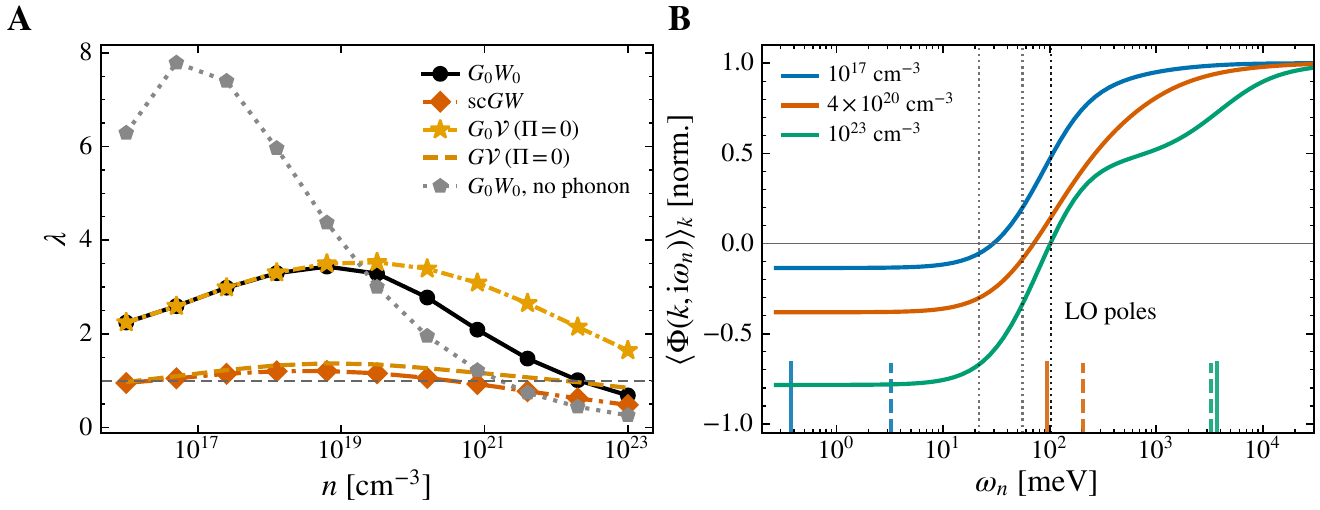}
\caption{(A)~Decomposition of the pairing kernel at $T=1$~K.  The plotted decompositions show $G_0\mathcal{V}$ and $G\mathcal{V}$, which remove the electronic polarization ($\Pi \equiv 0$) while keeping the lattice dielectric factor, and the one-shot no-phonon kernel, which sets $F_3(\nu) = 1$ and leaves a pure Coulomb interaction screened only electronically.  (B)~Momentum-averaged leading gap eigenfunction of the fully self-consistent sc$GW$ kernel versus Matsubara frequency at three densities (normalized to its high-frequency value).  The sign change sits within the polar-phonon scale window (nodes at $30$, $71$, and $102$~meV for the three densities; dotted lines mark the LO poles); short colored vertical ticks mark $E_{\mathrm{F}}$ (solid) and $\omega_{\mathrm{pl}}$ (dashed) for each density.
\label{fig:mechanism}}
\end{figure*}

We decompose the interaction in Fig.~\ref{fig:mechanism}(A) with three checks.

First, deleting the electronic polarization has a negligible effect on the low-density flank.  At the one-shot level, the strict $G_0\mathcal{V}$ kernel is indistinguishable from $G_0W_0$ for $n \lesssim 3\times10^{17}\,\mathrm{cm}^{-3}$ and agrees within a few percent at $6.3\times10^{18}\,\mathrm{cm}^{-3}$.  With the dressed $G$, the same variant changes the dilute eigenvalues only weakly.  In this low-density variant the pairing is therefore carried by the frequency structure of the lattice-screened Coulomb interaction, Eq.~\eqref{eq:Vtot}, rather than by the electronic-polarization channel of the present RPA kernel.  At high density the curves separate because electronic polarization screens the long-range interaction and closes the dome; at $10^{23}\,\mathrm{cm}^{-3}$ the one-shot eigenvalue drops from $1.65$ without electronic polarization to $0.69$ with it, and from $0.85$ to $0.48$ after dressing $G$.

Second, the no-phonon variant answers the converse question, and it is here that the one-shot kernel fails most visibly.  At the $G_0W_0$ level the pure-Coulomb kernel produces enormous eigenvalues, up to $\lambda = 7.8$ near $5\times10^{16}\,\mathrm{cm}^{-3}$, stronger than the full interaction.  This is the imaginary-axis counterpart of the RPA-level plasmon superconductivity found by Rietschel and Sham~\cite{RietschelSham83}, which Grabowski and Sham subsequently showed to be destroyed by vertex corrections~\cite{GrabowskiSham84}.  The dynamically screened Coulomb interaction alone therefore appears as a strong pairing channel only at the unrenormalized RPA level in this calculation; the no-phonon result should be read as a diagnostic of RPA charge-fluctuation pairing, not as a stable mechanism after normal-state renormalization and vertex corrections.

Third, the gap eigenfunction corroborates the mechanism.  Fig.~\ref{fig:mechanism}(B) shows the momentum-averaged leading sc$GW$ eigenfunction versus Matsubara frequency.  At every density the gap changes sign at a frequency set by the polar-phonon scale: the node sits at $30$, $71$, and $102$~meV for $n = 10^{17}$, $4\times10^{20}$, and $10^{23}\,\mathrm{cm}^{-3}$ respectively, inside the same LO pole window in which Fig.~\ref{fig:scales}(B) shows $F_3(\nu)$ rising from the statically screened limb toward the high-frequency Coulomb limb, while $E_{\mathrm{F}}$ and $\omega_{\mathrm{pl}}$ vary by four orders of magnitude over the same range.  This is the Bogoliubov-Tolmachev-Morel-Anderson sign structure~\cite{Bogoliubov59,MorelAnderson62} by which an everywhere-repulsive but retarded interaction \REV{leads to pairing}, with the retardation scale set by the polar phonon.  At the highest density ($10^{23}\,\mathrm{cm}^{-3}$), the three-mode lattice factor has already completed its LO-pole crossover before the eigenfunction develops its slower shoulder over $10^{3}$-$10^{4}$~meV.  This shoulder should therefore not be read as an additional polar-phonon scale; it occurs instead on the electronic scales $E_{\mathrm{TF}}\approx 2.14$~eV, $\omega_{\mathrm{pl}}\approx 3.26$~eV, and $E_{\mathrm{F}}\approx 3.74$~eV, reflecting the dynamical electronic screening entering $W$ and consistent with the separation of the decomposition curves in panel~(A).

We conclude that, within this model class, superconductivity in dilute STO is phonon-mediated in the precise sense that the pairing channel \REV{is provided by} the retarded frequency structure of the lattice-screened Coulomb interaction.  The kernel itself remains repulsive at every Matsubara frequency; it is the contrast between its lattice-screened low-frequency and Coulomb-dominated high-frequency limbs that drives the instability.  This conclusion does not rule out distinct plasmon-based models with different assumptions about screening in the dilute limit~\cite{RuhmanLee16}; it states that, for the present one-band RPA kernel (with vertex corrections excluded), the electronic charge-fluctuation channel is not the robust low-density mechanism, while electronic screening matters mainly by terminating the dome at high density.

\subsection{Fermi-surface projection, frequency truncation, and \texorpdfstring{$\mu^\ast$}{mu*}}\label{subsec:projection}

\begin{figure}[!htb]
\includegraphics[width=0.49\textwidth,clip]{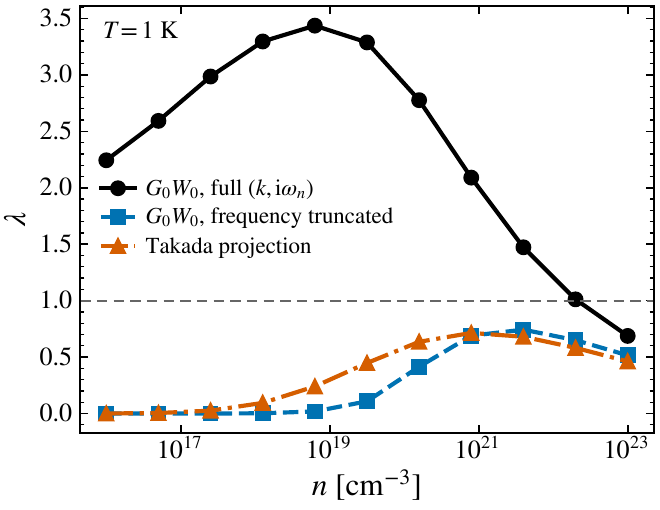}
\caption{Consequences of restricting the pairing kernel, all at the $G_0W_0$ level and $T = 1$~K.  Full kernel (black): all momenta and all Matsubara frequencies.  Truncated (blue): same kernel with the Matsubara window of the gap solve restricted to $|\omega_m| \leqslant E_{\mathrm{F}}$.  Takada projection (red):  Fermi-surface-projected solver following Ref.~\cite{Takada80-STO}.  Both restrictions eliminate the dilute-limit pairing.
\label{fig:projection}}
\end{figure}

Fig.~\ref{fig:projection} addresses question~(\lroman{4}) of the Introduction.  Restricting the Matsubara window of the gap equation to $|\omega_m| \leqslant E_{\mathrm{F}}$, the window implicitly assumed whenever the problem is reduced to a Fermi-surface theory, collapses the dilute eigenvalue from $\lambda \approx 2.2$-$3.4$ to zero for $n \lesssim 10^{19}\,\mathrm{cm}^{-3}$, with a Takada-style projection onto the Fermi surface~\cite{Takada80-STO} behaving in a similar manner.  The two restricted calculations disagree qualitatively with the full solution.  The reason is the scale separation of Fig.~\ref{fig:scales} run in reverse: when $E_{\mathrm{F}}$ lies below the polar LO pole window, the attraction-like frequency contrast of Eq.~\eqref{eq:Vtot} lives through the LO frequency range rather than inside the narrow electronic window, and, through the pair propagator, at momenta up to $k_{\mathrm{LO},j} = \sqrt{2m^\ast \omega_{\mathrm{LO},j}}$, which exceed $k_{\mathrm{F}}$ by orders of magnitude in the dilute limit.  A kernel restricted to the Fermi surface sees only the residual repulsion.  This is the numerical counterpart of the analytic observation of Gastiasoro, Chubukov, and Fernandes~\cite{Gastiasoro19-lowdensity} and Phan and Chubukov~\cite{PhanChubukov22-lowdensity} that the dilute pairing problem is controlled by states far from the Fermi surface.  At high density ($E_{\mathrm{F}}$ above the LO pole window) the three calculations converge, as they should.

This also explains why we do not introduce a Coulomb pseudopotential.  The $\mu^\ast$ construction integrates out the repulsion between a characteristic polar LO scale and $E_{\mathrm{F}}$, producing the one-scale reduction factor $1/[1 + \mu \ln(E_{\mathrm{F}}/\omega_{\mathrm{LO}})]$~\cite{MorelAnderson62}; on the low-density flank the hierarchy is inverted and there is no wide electronic window to integrate out.  Recent electron-liquid calculations likewise find that the phenomenological pseudopotential picture conflates Fermi-liquid renormalization with changes in the low-energy pairing vertex and is not a controlled substitute for microscopic Coulomb dynamics~\cite{Wang23-coulomb-pseudopotential,Cai25-electron-liquids}.  The dynamics that $\mu^\ast$ is meant to summarize is precisely the dynamics the full kernel must keep.  In our calculations the entire Tolmachev-Morel-Anderson physics is contained in the explicit frequency dependence of $W$ and in the sign change of the gap function [Fig.~\ref{fig:mechanism}(B)].

Two caveats belong with this comparison.  The full-kernel solution is more physical than the projected or truncated one, but it is still an RPA-level theory: the lesson of Refs.~\cite{RietschelSham83,GrabowskiSham84} is that kernels of this class overestimate pairing from dynamical charge fluctuations unless vertex corrections are included.  Recent quasiparticle $GW$ work for superconductors found a spurious plasmon-mediated pairing enhancement at the s-$GW$ level in graphene and Nb and showed that a quasiparticle projection suppresses this artifact; in that construction the screened interaction is held at its initial RPA value~\cite{Spataru26-sqpGW}, unlike the present sc$GW$ update of $W$ from a polarization built with dressed Green functions.  The normal-state self-consistency shown in Sec.~\ref{subsec:mechanism} suppresses the same tendency, but it is not a substitute for the inclusion of an explicit vertex function.

\subsection{Pairing-onset dome and the density limits}\label{subsec:tc-limits}

\begin{figure}[!htb]
\includegraphics[width=0.49\textwidth,clip]{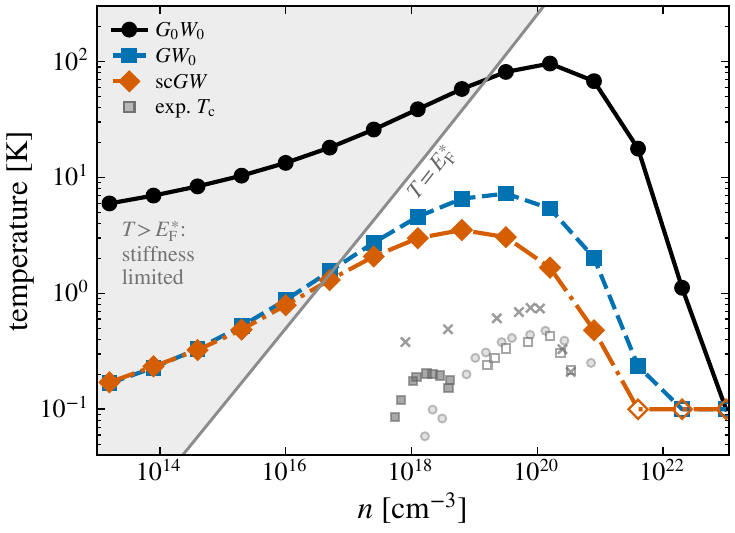}
\caption{Pairing-onset temperature from the crossing $\lambda(T_{\mathrm{c}}) = 1$ for the three kernels, together with experimental bulk STO and STO-based transition temperatures.  Open colored symbols denote calculated densities at which the crossing lies below the sampled window ($T < 0.1$~K).  \ZHC{The gray line is $T = E_{\mathrm{F}}^\ast$, estimated from $E_{\mathrm{F}}/Z(k_{\mathrm{F}},\ii0)$ using a low-frequency extrapolation of the sc$GW$ normal-state solution:} in the shaded region the computed onset exceeds the degeneracy scale and the physical transition is bounded by the phase stiffness, \ZHC{$T_\theta \sim E_{\mathrm{F}}^\ast$}.  Experimental symbols are filled gray squares for oxygen-reduced samples, open gray squares for Nb-doped samples, and gray circles for Schooley et al.\ comparison data, digitized from Ref.~\cite{Lin14,Schooley64}; gray crosses show Nb:Sr$_{0.95}$Ba$_{0.05}$TiO$_3$ from Tomioka et al.~\cite{Tomioka22-polar-metal}.
\label{fig:tc}}
\end{figure}

Fig.~\ref{fig:tc} converts the eigenvalue curves into onset temperatures.  The hierarchy of Sec.~\ref{subsec:selfconsistency} is amplified by the sensitivity of $T_{\mathrm{c}}$.  The $G_0W_0$ kernel peaks at $T_{\mathrm{c}} \approx 96$~K near $1.6\times10^{20}\,\mathrm{cm}^{-3}$; dressing the propagator ($GW_0$) lowers the peak to $\approx 7.3$~K near $3.2\times10^{19}\,\mathrm{cm}^{-3}$, and the screening update of sc$GW$ lowers it again to $\approx 3.5$~K near $6.3\times10^{18}\,\mathrm{cm}^{-3}$.  The compressed one-mode reference ($\omega_{\mathrm{LO}}^{\mathrm{eff}} \sim 84$ meV) gives the same hierarchy, with a sc$GW$ peak of $3.93$~K.  The sc$GW$ suppression grows toward high density: for instance, at $4.0\times10^{21}\,\mathrm{cm}^{-3}$ the sc$GW$ crossing falls below the $0.1$~K window while $GW_0$ still pairs weakly.  Each stage of self-consistency therefore lowers the dome and narrows it from the high-density side.

The two flanks of the dome have different explanations, which we make quantitative in Appendix~\ref{app:limits} for distinct density limits.  On the high-density side, where $E_{\mathrm{F}}$ exceeds the LO pole window, projecting the interaction on the Fermi surface is legitimate.  In the one-scale notation used for the analytic estimates in Appendix~\ref{app:limits}, the Fermi-surface average of the long-range kernel gives an effective coupling $\eta \propto \ln(2k_{\mathrm{F}}/q_{\mathrm{TF}})/k_{\mathrm{F}}$ that decreases with density; a two-window Bogoliubov-Tolmachev analysis then gives $T_{\mathrm{c}} \sim \omega_0\, \ee^{-1/\eta}$, crossing over at the highest densities to $T_{\mathrm{c}} \sim \omega_0\, \exp[-1/(\eta^2 \ln(E_{\mathrm{F}}/\omega_0))]$ once the retardation window opens.  The steep collapse of the computed curves is consistent with this exponential sensitivity, with the additional electronic screening of Sec.~\ref{subsec:mechanism} accelerating it.

On the low-density side, the same Fermi-surface logic would predict an exponentially small onset: in the one-scale notation the projected coupling is $\bar\eta = \eta\, E_{\mathrm{F}}^2/2\omega_0^2 \propto n$, and $T_{\mathrm{c}}^{\mathrm{FS}} \sim E_{\mathrm{F}}\sqrt{\bar\eta}\, \ee^{-1/\bar\eta^2}$ (Appendix~\ref{app:limits}), i.e.\ unobservably small below $10^{19}\,\mathrm{cm}^{-3}$.  The full calculation instead remains non-zero on the sampled dilute grid and tends toward the density-independent dilute kernel of the full momentum-frequency gap equation~\cite{Gastiasoro19-lowdensity,PhanChubukov22-lowdensity}: $GW_0$ and sc$GW$ give pairing onsets of $\approx 0.88$ and $0.79$~K at $10^{16}\,\mathrm{cm}^{-3}$ and nearly coincide at $0.17$~K by $1.6\times10^{13}\,\mathrm{cm}^{-3}$.  The computed dilute onsets are of the same order as the $\mu \to 0$ onset found in Ref.~\cite{PhanChubukov22-lowdensity} for comparable single-scale parameters.

The dilute onset, however, is not the measured $T_{\mathrm{c}}$.  \ZHC{Below $n \approx 3\times10^{16}\,\mathrm{cm}^{-3}$ the computed onset exceeds the renormalized Fermi scale ($E_{\mathrm{F}}^\ast \approx 0.50$~K at $10^{16}\,\mathrm{cm}^{-3}$), and the transition is necessarily bounded by the phase stiffness, $T_\theta \sim E_{\mathrm{F}}^\ast$}~\cite{EmeryKivelson95}; the measured superfluid density in STO indeed tracks the carrier density on the underdoped side~\cite{Collignon17,Thiemann18}.  In the shaded region of Fig.~\ref{fig:tc} the $\lambda = 1$ crossing should be read as the onset of local pairing~\cite{Eagles69}, with phase-coherent superconductivity (or a BEC of preformed pairs) appearing only at the lower scale $T_\theta$.  The sc$GW$ dome has roughly the experimental shape, with its maximum at $6.3\times10^{18}\,\mathrm{cm}^{-3}$ lying about an order of magnitude below the measured peak position near $1$-$2\times10^{20}\,\mathrm{cm}^{-3}$, and its peak onset of $\approx 3.5$~K exceeding the measured $T_{\mathrm{c}} \approx 0.4$~K by a factor of ten.  This residual mismatch is the quantitative result of the vertex-free, single-band, isotropic theory.

\section{Discussion}\label{sec:discussion}

\REV{We have used a full numerical solution of the $GW$ equations to study superconductivity and normal-state properties of lightly doped, nearly ferroelectric \ce{SrTiO3} models, in which carriers interact through the Coulomb interaction and couple to polar phonons in the presence of the very strong and strongly frequency-dependent screening arising from incipient ferroelectricity. We distinguish high-density (Fermi energy and plasma frequency high compared with the LO phonon scale) and low-density regimes. In the high-density regime the usual considerations of superconductivity in a Fermi liquid apply; in the low-density regime the system is a gas of quasiparticles strongly renormalized (by a factor of $\sim 2$) by phonons, with interactions important only at frequencies well above the Fermi energy.  This situation presents unusual physics: the conventional arguments leading to Wigner crystal or small-polaron formation do not apply because the static interaction is vanishingly small, but high-frequency interactions lead to relatively strong pairing tendencies. The calculations that lead to these results, however, rely on several approximations, both to the mathematics and to the physics. In this section we discuss the approximations in detail.}

\emph{Vertex corrections.}  The leading diagrams absent from our kernels are the vertex corrections~\cite{McClain16}, and in the antiadiabatic regime nothing protects their neglect: the relevant ratios $\omega_{\mathrm{LO},j}/E_{\mathrm{F}}$ are large, and the long-range polar coupling is intermediate rather than weak~\cite{Chubukov20-eliashberg}.  Recent analyses of electron-phonon models also emphasize that Migdal-Eliashberg theory~\cite{Migdal58,Eliashberg60} can fail through polaronic or bipolaronic physics before a soft-phonon instability is reached~\cite{Chubukov26-MET-breakdown}.  What is known suggests that vertex corrections are likely to suppress the present RPA-level onsets, but the sign and size are not settled.  For the charge-fluctuation (plasmon) channel the classic result of Grabowski and Sham~\cite{GrabowskiSham84} is that low-order vertex corrections drastically weaken RPA-level pairing; our finding that self-consistency removes most of the one-shot plasmon pairing [Fig.~\ref{fig:mechanism}(A)] points the same way, but explicit vertices are needed to test how much remains.  For the phonon channel the sign is less obvious: the nonadiabatic perturbation theory of Pietronero, Str\"assler, and Grimaldi~\cite{Grimaldi95,Pietronero95} finds that vertex corrections \emph{enhance} pairing when the coupling is dominated by small momentum transfers, as is the case for $1/q$ Fr\"ohlich coupling, while for the dilute polar model Gastiasoro \etal~\cite{Gastiasoro19-lowdensity} identified the failure of the Fermi-surface restriction as the dominant correction in the dilute regime and argued that the remaining corrections, including vertices, are likely to be modest; explicit vertex corrections were not calculated.  A simple static screening estimate along the lines of a ladder-dressed coupling, $\tilde g_q = g_q/[1 + |g_q|^2 |D_0(q,0)\,\Pi_0(q,0)|]$, gives $|\tilde g| < |g|$ and hence additional suppression of the polar channel. \REV{The near vanishing of the interaction at the Fermi level means that the situation is somewhat different from the situations considered in most published vertex correction calculations.} Settling the sign and size of the vertex contribution for this model, by a $GW\Gamma$ extension~\cite{Kutepov16-Hedin-vertex} or by diagrammatic Monte Carlo, is in our view the most important open theoretical problem in this system, and the present vertex-free results provide the reference point such a calculation needs.

\emph{Phase fluctuations.}  \REV{The low-density side of the superconducting dome can be further separated into two regimes: moderately low densities, in which the pairing scale remains small compared to the Fermi energy, and very low densities, where the pairing scale becomes comparable to the Fermi energy. In this second regime, one is dealing with a three-dimensional Bose condensate situation, and the actual transition will be determined by phase fluctuations.} The stiffness bound of Fig.~\ref{fig:tc} is the only quantitative statement we can make beyond the pairing onset.  Converting the onset dome into a transition dome requires the superfluid stiffness and the fluctuation thermodynamics of a dilute pair fluid, which our linearized framework does not provide.  The experimental observation that the superfluid density equals the carrier density on the underdoped side~\cite{Collignon17} suggests that the measured transition tracks the stiffness scale well before the dilute onset of Fig.~\ref{fig:tc} could be reached.

\emph{Multiband structure, anisotropy, and disorder.}   \REV{In \ce{SrTiO3}} the conduction band is three $t_{2g}$ sheets split by spin-orbit coupling and the tetragonal distortion~\cite{Marel11,Lin14}, with anisotropic masses.  Below the first Lifshitz transition only the lowest band is occupied, but the experimental dome maximum sits near the densities where the second and third bands fill, and interband processes plausibly reshape the peak~\cite{Binnig80}; microwave data in Nb-doped STO suggest that weak disorder can homogenize the gap across occupied bands~\cite{Thiemann18}.  Anisotropy modifies the angular average of the $1/q^2$ kernel and should change $\lambda$ at the tens-of-percent level but not qualitatively.  Disorder, oxygen vacancies in particular, affects both the carriers and the polar response and remains a sample-dependent complication~\cite{Collignon19}. \REV{Our numerical methods, which rely on simplifications associated with the spherical approximation of a single parabolic band, are not yet able to deal with {disorder, oxygen vacancies and similar effects, which become increasingly important at very low carrier densities.}  Our results should be regarded as {results of an idealized model rather than as a quantitatively predictive description of all samples. }}  Finally, our model contains the soft TO sector only through the longitudinal lattice dielectric response and the static dielectric constant $\epsilon_0$.  It does not include an independent transverse soft-mode propagator~\cite{Zhou18-STO-soft-modes,Huang25-NbOI2-softTO}, spin-orbit-assisted coupling to polar distortions~\cite{Gastiasoro23-Rashba}, nonlinear couplings in the ferroelectric ordered phase, or the doping dependence of the polar response~\cite{Fauque25-dipolar-length}. \ZHC{In Appendix~\ref{app:fauque-softmode} we explore the effects of the doping dependence of the polar response in a minimal model context, finding only a modest change of the estimated transition temperatures which occur only in the high-density regime.} The microscopic coupling of dilute carriers to the soft TO mode remains debated~\cite{Wolfle18-STO,RuhmanLee19-comment,WolfleBalatsky19-reply}.  The possibility of a net attractive static long-range interaction from multiple LO modes has likewise been debated~\cite{Tubman25-STO,Ruhman25-multimode-comment,Tubman26-downfolding}.  Proposals in which the ferroelectric soft mode couples directly to the carriers~\cite{Edge15,Marel19,Volkov22,Yu22-quantum-paraelectrics,Saha25-strong-coupling,Rischau17-ferroelectric,Rischau22-isotope,Yoon21} involve physics outside the interaction~\eqref{eq:Vtot}; approaches emphasizing phonon-induced attractive effective interactions~\cite{Ngai74,KiselovFeigelman21} are likewise complementary to, rather than excluded by, our results.

\section{Concluding remarks}\label{sec:conclusion}

We have presented numerically converged finite-temperature solutions of the self-consistent $GW$ (renormalized Migdal-Eliashberg) equations for polar one-band models of doped \ce{SrTiO3}.  The calculation treats the Coulomb repulsion and the three polar-phonon branches through a dynamically screened interaction and solves the full momentum- and frequency-resolved gap equation without truncations.

Normal-state self-consistency is central in this problem.  The Eliashberg renormalization $Z \approx 1.9$ of the dilute polaronic carriers reduces the pairing eigenvalue by a factor of roughly $2$-$3$ and the onset temperature by one to two orders of magnitude relative to one-shot $G_0W_0$; the screening update of sc$GW$ then halves the dome maximum and suppresses the high-density flank.  Decomposition calculations identify the retarded structure of the lattice-screened Coulomb interaction as the low-density pairing channel and electronic screening as the agent that closes the high-density side, while exposing strong one-shot pure-Coulomb pairing as an RPA artifact.  The full equation retains a non-zero dilute-limit pairing-onset scale, carried by states far from the Fermi surface; Fermi-surface projection or truncation eliminates it.  The computed onset still exceeds the measured transition scale by about an order of magnitude, leaving vertex corrections, multiband effects, and phase fluctuations as the leading missing ingredients.

Beyond the specific material, these results provide detailed numerical evidence for how superconductivity emerges in dilute polar materials, and they carry a methodological message: quantitative statements about pairing in this regime require self-consistency and the full momentum and frequency structure of the kernel, and will ultimately require vertex corrections.  A similar need for material-specific, quantitatively controlled many-body treatments appears in \abinitio studies of cuprate electronic structure and superconducting trends~\cite{Cui21-cuprate-parent-state,Cui23-cuprate-doping}.  The same considerations apply to other dilute polar superconductors, such as Tl-doped PbTe~\cite{Matsushita06} and the KTaO$_3$-based electron gases~\cite{Liu21-KTO}.  Beyond this class, related first-principles $GW$ treatments of electron-phonon superconductivity in bismuthates and other correlated superconductors likewise show that many-body corrections can materially affect pairing estimates~\cite{Li19-BKBO-GWPT,Li24-Nickelate-GWPT,You25-Kagome-GWEliashberg}.  The methodology itself (radial reduction with analytic treatment of the Coulomb singularity, sparse-IR frequency handling, and a Davidson gap solver, all released in the open-source package \textsc{migdal}~\cite{MigdalCode}) extends directly to multiband and anisotropic generalizations, and to vertex-corrected kernels, which we leave for future work.

\begin{acknowledgments}
We thank Yao Luo, Jin-Jian Zhou, Tianyu Zhu, Maria Gastiasoro, and Shu Fay Ung for helpful discussions. ZHC, AJM, and DRR were funded by the
Columbia MRSEC on Precision-Assembled Quantum Materials (PAQM) under award number
DMR-2011738. We also acknowledge a generous donation from the Keele Foundation.  The Flatiron
Institute is a division of the Simons Foundation.

The \textsc{migdal} package and runnable scripts are available at
	\url{https://github.com/zhcui/migdal_preview}.
\end{acknowledgments}

\appendix

\section{Ab initio parametrization of one-band models}\label{app:abinitio}

\begin{figure*}[!htb]
\includegraphics[width=0.98\textwidth,  clip]{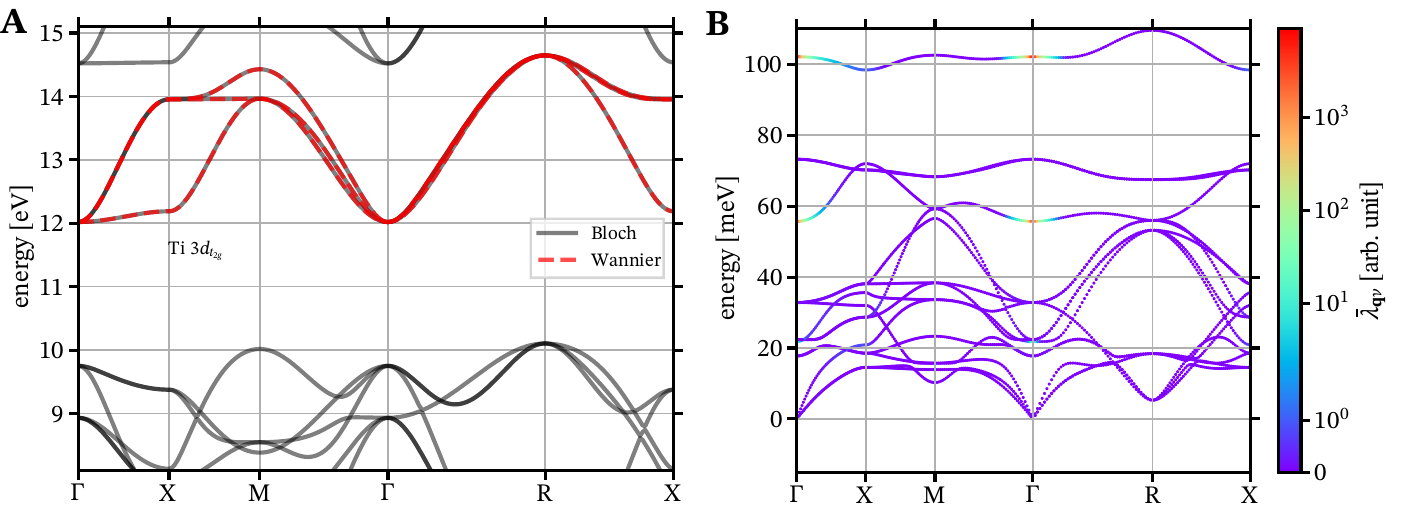}
\caption{(A) Electronic band structure of cubic \ce{SrTiO3}. The lowest three conduction bands (with Ti $t_{2g}$ orbital character) are localized as Wannier functions. (B) Phonon dispersion of \ce{SrTiO3} calculated using DFPT+TDEP. The color map indicates the normalized mode-resolved electron-phonon coupling parameter $\bar{\lambda}^{\mathrm{ep}}_{\mathbf{q}\nu}$.
\label{fig:dft}}
\end{figure*}

The density functional theory (DFT) and DFPT calculations were performed with \textsc{Quantum ESPRESSO}~\cite{QEcode,QE17,Baroni01}, using the PBE generalized-gradient approximation~\cite{Perdew96PBE} and norm-conserving UPF pseudopotentials from the ONCVPSP/PseudoDojo family~\cite{Hamann13-ONCVPSP,PseudoDojo18}.  The cubic five-atom \ce{SrTiO3} cell used $a=3.90$~\AA, a plane-wave kinetic-energy cutoff of $85$~Ry, an unshifted $16\times16\times16$ $k$ mesh, and an electronic self-consistency threshold of $10^{-12}$~Ry.  DFPT perturbations were computed on an $8\times8\times8$ $q$ mesh, and the TDEP force constants used here were fitted from 300~K samples in a $3\times3\times3$ supercell.

Fig.~\ref{fig:dft}(A) shows the DFT band structure of cubic \ce{SrTiO3} together with the maximally localized Wannier interpolation of the three Ti $t_{2g}$ conduction bands (using \textsc{wannier90}~\cite{Marzari12RMP,Mostofi14}).  The $t_{2g}$ bands are anisotropic, with $m^\ast$ ranging from $0.5$ to $6\,m_{\mathrm{e}}$ along $\Gamma$-$X$ and $0.5$ to $0.8\,m_{\mathrm{e}}$ along $\Gamma$-$R$.  At the conduction-band minimum the three $\Gamma$-centered valleys are symmetry-equivalent (without spin-orbit coupling); using one heavy mass $m_{\mathrm{h}}$ and two light masses $m_{\mathrm{l}}$ along $\Gamma\!\to\!X$, the density-of-states mass is
\begin{equation}
  m^\ast_{\mathrm{DOS}} = g_{\mathrm{v}}^{2/3}\,(m_{\mathrm{h}} m_{\mathrm{l}}^2)^{1/3} \sim 2.1\, m_{\mathrm{e}},
  \qquad g_{\mathrm{v}}=3,
\end{equation}
which reconciles the large single-direction heavy mass with the $\sim 2\, m_{\mathrm{e}}$ values quoted from DOS- and transport-sensitive measurements.  In principle one should treat the low-temperature tetragonal phase with spin-orbit coupling; for the one-band model of the main text we use the averaged $m^\ast = 2.1\,m_{\mathrm{e}}$ and have checked the robustness of the results within $0.5$-$4\,m_{\mathrm{e}}$.

Fig.~\ref{fig:dft}(B) shows the phonon dispersion together with a color map of the normalized mode-resolved electron-phonon coupling parameter $\bar{\lambda}^{\mathrm{ep}}_{\mathbf{q}\nu}$.  The underlying unnormalized quantity is computed from the standard phonon-linewidth expression
\begin{equation}
\begin{split}
  \lambda^{\mathrm{ep}}_{\mathbf{q}\nu}
  =
  \frac{1}{N(\varepsilon_{\mathrm{F}})\omega_{\mathbf{q}\nu}}
  \sum_{mn}\int_{\mathrm{BZ}} &\frac{\diff\mathbf{k}}{\Omega_{\mathrm{BZ}}}
  |g_{mn\nu}(\mathbf{k},\mathbf{q})|^2\, \times \\
  &\delta(\varepsilon_{n\mathbf{k}}-\varepsilon_{\mathrm{F}})
  \delta(\varepsilon_{m,\mathbf{k}+\mathbf{q}}-\varepsilon_{\mathrm{F}}).
\end{split}
\end{equation}
The finite-temperature force constants were obtained from the TDEP construction~\cite{Hellman11-TDEP,Hellman13-TDEP,Knoop24-TDEP}, and the long-range polar electron-phonon matrix elements were extracted with \textsc{Perturbo}~\cite{Perturbo21}.  The dominant weight is on the highest LO polar branch near the zone center, consistent with the strong small-$q$ enhancement of the underlying Fr\"ohlich matrix element; all other branches couple weakly on the relevant energy and momentum scales.  This justifies the reduction to a long-range polar interaction, the long-range part of Eq.~\eqref{eq:Vtot}.

For the dielectric function we use the three-pole factor of Eq.~\eqref{eq:Vtot}.  The real material has several infrared-active polar doublets. In the charge-neutralized 300~K cubic \abinitio/TDEP parametrization used here, the soft TO response carries essentially the entire static enhancement, while the LO zeros span the polar crossover window.  The anchored pole set used in the main text is
\begin{equation}
\begin{split}
  \omega_{\mathrm{TO}} &= (1.24,\;22.39,\;73.24)~\mathrm{meV},
\\
  \omega_{\mathrm{LO}} &= (21.80,\;55.72,\;102.14)~\mathrm{meV}.
\end{split}
\end{equation}
The higher TO poles and all LO zeros are obtained from the DFPT/TDEP
non-analytic $\Gamma$ limit.  The lowest TO pole is then replaced by the value
required by the experimental low-temperature static dielectric constant, giving
\begin{equation} \prod_{j=1}^{3}\frac{\omega_{\mathrm{LO},j}^{2}}{\omega_{\mathrm{TO},j}^{2}}
 = \frac{\epsilon_0}{\epsilon_\infty} = 3731.90.
\end{equation}
Thus, with the higher TO poles and all LO zeros held fixed, $\omega_{\mathrm{TO},1}=1.24$~meV is determined by $\epsilon_0$ through the generalized LST relation rather than taken from a direct spectroscopic soft-mode measurement.
 This construction preserves both $F_3(0)=\epsilon_\infty/\epsilon_0$ and $F_3(\infty)=1$ while retaining the frequency structure of the three polar branches.

For comparison only, we also generated a one-mode compressed data set using the log-strength average defined in the main text, giving $\omega_{\mathrm{LO,eff}}=84.26$~meV and the LST-consistent $\omega_{\mathrm{TO,eff}}=1.38$~meV.  This reference is useful as a robustness check, but it is not the source of the main-text figures.

\section{Connection between self-consistent \texorpdfstring{$GW$}{GW} and renormalized Migdal-Eliashberg theory}\label{app:equivalence}

In a coupled electron-phonon treatment the phonon Green's function is
\begin{equation}
D(\vecq, \ii\nu_n) = -\frac{1}{\nu^2_n + \omega_{\vecq}^2 + \Pi^{\mathrm{ph}} (\vecq, \ii \nu_n)},
\label{eq:phonon GF}
\end{equation}
and the Migdal electron-phonon self-energy and phonon self-energy are
\begin{align}
\Sigma^{\mathrm{el\text{-}ph}}(\veck, \ii \omega_m)
&= -\frac{1}{\beta}  \sum_{n} \int_{\vecq}
\abs{g_{\vecq}}^2 \,
G(\veck - \vecq, \ii \omega_m - \ii\nu_n)
\nonumber\\
&\quad \times D(\vecq, \ii\nu_n),
\label{eq:migdal el-ph self energy}\\
    \Pi^{\mathrm{ph}} (\vecq, \ii \nu_n) &= \abs{g_{\vecq}}^2\,\Pi(\vecq,\ii\nu_n),
\label{eq:phonon self energy}
\end{align}
with $\Pi$ the $GG$ bubble of Eq.~\eqref{eq:Pi}.  In Marsiglio's terminology~\cite{Marsiglio90-migdal-renorm}, the \emph{unrenormalized} Migdal-Eliashberg theory fixes $D = D_0$ (no phonon self-energy), while the \emph{renormalized} theory keeps Eq.~\eqref{eq:phonon self energy}.

Consider first a system with no direct Coulomb term and take $\Pi = 0$ in Eq.~\eqref{eq:effective W}.  Then $W = \mathcal{V} = |g_q|^2 D_0$ and Eq.~\eqref{eq:effective GW} reduces to Eq.~\eqref{eq:migdal el-ph self energy} with $D \to D_0$: the unrenormalized theory is the unscreened (self-consistent Hartree-Fock-like) limit of the $GW$ form.  With nonzero $\Pi$,
\begin{equation}
\begin{split}
    \Sigma(\veck, \ii \omega_m)
    &= -\frac{1}{\beta} \sum_{n} \int_{\vecq}
    G(\veck - \vecq, \ii \omega_m - \ii \nu_n)
    \\
    &\quad \times
    \frac{\abs{g_{\vecq}}^2 D_0}
         {1 - \Pi\, \abs{g_{\vecq}}^2 D_0}
    \\
    &= -\frac{1}{\beta}  \sum_{n} \int_{\vecq}
    G(\veck - \vecq, \ii \omega_m - \ii \nu_n)
    \\
    &\quad \times
    \abs{g_{\vecq}}^2 D(\vecq, \ii \nu_n),
\end{split}
\end{equation}
where the second line uses Eqs.~\eqref{eq:phonon GF} and \eqref{eq:phonon self energy}.  The RPA denominator of the $GW$ screened interaction is therefore the same Dyson resummation as the phonon renormalization in Marsiglio's formulation: self-consistent $GW$ with the interaction $\mathcal{V}$ of Eq.~\eqref{eq:Vtot} is the bare-vertex, phonon-renormalized Migdal-Eliashberg approximation, written so that the Coulomb repulsion and the electron-phonon attraction are screened by the same electronic polarization rather than treated separately. 

The linearized anomalous self-energy in this framework is Eq.~\eqref{eq:LGE} of the main text.  In the isotropic single-band, degenerate limit, one may project Eq.~\eqref{eq:LGE} onto the Fermi surface and recover the familiar two-equation Eliashberg form for $Z(\ii\omega_m)$ and $\Delta(\ii\omega_m)$ with a Fermi-surface-averaged kernel and a Morel-Anderson pseudopotential; Sec.~\ref{subsec:projection} of the main text discusses when this projection fails.

\section{Numerical implementation}\label{app:implementation}

Our code exploits spherical symmetry, uses sparse-IR transformations for imaginary time and frequency, and treats the Coulomb head of the radial kernel analytically rather than by pointwise interpolation.  We use PySCF~\cite{Sun20pyscf} for the Davidson eigensolver and DIIS mixing.  The implementation, including all drivers used in this paper, is distributed as the open-source package \textsc{migdal}~\cite{MigdalCode}.

\paragraph*{Radial reduction.}
For any isotropic convolution, the angular variables may be integrated out by trading the scattering angle for the momentum transfer $q=|\veck-\vecp|$.  With $\int_{\mathbf p}=\int \diff^3p/(2\pi)^3$, one has, for $k>0$,
\begin{equation}
\begin{split}
\int_{\mathbf p} F(p)\, W(|\veck-\mathbf{p}|)
&=\frac{1}{(2\pi)^2 k}\int_0^\infty \diff p\,p F(p)
\\
&\quad \times
\int_{|k-p|}^{k+p}\diff q\,q W(q).
\end{split}
\label{eq:radial-convolution}
\end{equation}
Equivalently,
\begin{equation}
\bar W(k,p;\ii\nu_n)=\frac{1}{kp}\int_{|k-p|}^{k+p}\diff q\,q W(q,\ii\nu_n),
\label{eq:Wbar-impl}
\end{equation}
with the $k\to0$ and $p\to0$ rows obtained from the analytic limit of Eq.~\eqref{eq:radial-convolution}.  In practice the code does not perform an angular quadrature for every matrix-vector product; it precomputes the primitives
\begin{equation}
I_W(q;\ii\nu_n)=\int_0^q \diff q'\, q' W(q',\ii\nu_n),
\end{equation}
so that the radial angular factor is simply $I_W(k+p)-I_W(|k-p|)$.

\paragraph*{Coulomb-head treatment.}
The radial integral is numerically delicate because the small-$q$ interaction contains a Coulomb head.  For each bosonic frequency the screened interaction is decomposed as
\begin{equation}
W(q,\ii\nu_n)=\frac{A_W(\ii\nu_n)}{q^2}+W_{\mathrm{reg}}(q,\ii\nu_n),
\label{eq:W-head-split}
\end{equation}
so that
\begin{equation}
\int_a^b \diff q\,q W(q,\ii\nu_n)
=A_W(\ii\nu_n)\ln\frac{b}{a}
+\int_a^b \diff q\, q W_{\mathrm{reg}}(q,\ii\nu_n)
\label{eq:q-singular-primitive}
\end{equation}
for $a>0$.  The regular part $qW_{\mathrm{reg}}$ is linearly interpolated on a nonuniform $q$ grid and integrated exactly on each segment.  The diagonal block $a=|k-p|=0$ is not evaluated as a point value; it is replaced by a finite-range radial-cell average of the logarithmic kernel, which removes the grid-dependent singular self-block without changing the continuum interaction (see the inset of Fig.~\ref{fig:conv-GW}(A) for the performance of cell average).

For a screened interaction the coefficient $A_W$ is inferred from the low-$q$ expansion after the RPA denominator has been applied.  Writing
\begin{equation}
\begin{aligned}
\mathcal{V}(q,\ii\nu_n)
&=\frac{B(\ii\nu_n)}{q^2}
  +\mathcal{V}_{\mathrm{reg}}(q,\ii\nu_n),
\\
\Pi(q,\ii\nu_n)
&=P_0(\ii\nu_n)+C(\ii\nu_n)q^2+O(q^4),
\end{aligned}
\end{equation}
the static metallic response with $P_0\ne0$ screens away the $1/q^2$ tail, giving $A_W=0$.  For non-zero bosonic frequencies, charge conservation gives $P_0=0$~\cite{VanHoucke17-jellium-GW}, and the remaining singular coefficient is
\begin{equation}
A_W(\ii\nu_n)=\frac{B(\ii\nu_n)}{1-C(\ii\nu_n)B(\ii\nu_n)}.
\label{eq:screened-head}
\end{equation}

\paragraph*{Sparse-IR formulation.}
The intermediate representation basis~\cite{Shinaoka17,Li20-sparse-grids-gw-gf} provides compact representations of Green's functions, screened interactions, self-energies, and polarizations on the Matsubara and imaginary-time axes.  The IR basis and sparse-sampling transforms are generated with the \texttt{sparse-ir} library~\cite{Wallerberger23SparseIR}.  The basis follows from the singular-value decomposition of the finite-temperature spectral kernel,
\begin{equation}
K^{\alpha}(\tau,\omega)
=\sum_{\ell}U_\ell^{\alpha}(\tau)S_\ell^{\alpha}V_\ell^{\alpha}(\omega),
\qquad \alpha = \mathrm{F}, \mathrm{B},
\label{eq:IR-SVD}
\end{equation}
and sparse sampling in $\tau$ and $\ii\omega_n$ transforms between sampled values and IR coefficients.  Convolutions in Matsubara frequency become products in imaginary time, which is what makes the untruncated frequency treatment of the main text affordable at sub-kelvin temperatures.  We separate the instantaneous Coulomb part and the retarded remainder,
\begin{equation}
W(k,p;\ii\nu_n)=\bar v_{\mathrm{C}}(k,p)+\delta\bar W(k,p;\ii\nu_n);
\end{equation}
the static exchange is evaluated directly from the Coulomb primitive, while the dynamic self-energy is built in imaginary time.  For the scaled object $\widetilde\Sigma(k,\tau)=k\Sigma(k,\tau)$,
\begin{equation}
\begin{split}
\widetilde\Sigma_{\mathrm{dyn}}(k,\tau)
&= -\frac{1}{(2\pi)^2}
\int_0^\infty\diff p\,p\,G(p,\tau)
\\
&\quad \times
\left[I_{\delta W}(k+p,\tau)-I_{\delta W}(|k-p|,\tau)\right],
\end{split}
\label{eq:sigma-dyn-scaled}
\end{equation}
followed by an IR transform back to Matsubara frequency.  The self-consistent polarization of sc$GW$ is evaluated in the same radial geometry,
\begin{equation}
\begin{split}
\Pi(q,\tau)
&=-\frac{1}{2\pi^2 q}\int_0^\infty \diff k\,k\,G(k,\tau)
\\
&\quad \times
\int_{|k-q|}^{k+q}\diff p\,p\,G(p,\beta-\tau),
\end{split}
\label{eq:Pi-radial-tau}
\end{equation}
with an analytic $q\to0$ limit and the non-zero-frequency low-$q$ rows continued with the charge-conserving $\Pi \propto q^2$ form.

 After each update of $G$, the chemical potential is refit so that
\begin{equation}
n=\frac{1}{\pi^2}\int_0^\infty \diff k\,k^2 \rho(k)
\label{eq:number-eq-code}
\end{equation}
matches the target density; in the interacting calculations the particle number uses a high-frequency-corrected endpoint formula so the sparse-IR transform acts only on $G-G_0$.  The self-energy is damped and mixed by DIIS~\cite{Pokhilko22-Dyson-DIIS}, with the residual defined for the gauge-invariant combination $\Sigma-\mu$. These numerical techniques are particularly important for the challenging convergence in the low-density regime.

\paragraph*{Gap solver and density constraint.}
The instability is solved as a radial eigenvalue problem with the same convolution kernel.  To remove the removable $1/k$ singularity of Eq.~\eqref{eq:radial-convolution}, the Davidson vectors store the scaled gap variable $\widetilde\Phi(k,\ii\omega_n)=k\,\Phi(k,\ii\omega_n)$; the source entering the radial convolution is
\begin{equation}
S(k,\ii\omega_n)=G(k,\ii\omega_n)\,G(k,-\ii\omega_n)\,\widetilde\Phi(k,\ii\omega_n),
\end{equation}
and the matrix-vector product is the negative of the scaled anomalous self-energy convolution. 

\paragraph*{Settings used in this paper.}
Unless noted otherwise: $T = 1$~K for the eigenvalue scans, IR cutoff $\Lambda = 10^9$, $N_q = 500$ radial $q$ points, $N_k \approx 300$-$570$ nonuniform radial $k$ points adapted to $k_{\mathrm{F}}$ and $k_{\mathrm{LO}}$, Davidson tolerance $10^{-6}$, and SCF tolerance $10^{-6}$ with damping $0.98$ and DIIS for sc$GW$.  The $T_{\mathrm{c}}$ scans of Fig.~\ref{fig:tc} locate $\lambda(T_{\mathrm{c}})=1$ by cooling scans with interpolation; for this low-temperature scan we use $N_q=600$ (and $N_q=750$ for the lowest two densities), and each sc$GW$ point is initialized from the converged $GW_0$ solution at the same density and temperature. For the two lowest-density sc$GW$ chains we also use a minimum of 200 radial $k$ shells and SCF damping 0.995. The low-temperature endpoint of these scans uses the same $\Lambda=10^9$ frequency cutoff, which is necessary and is validated in Appendix~\ref{app:conv}.

\section{High- and low-density limits}\label{app:limits}

This appendix records the asymptotic analysis quoted in Sec.~\ref{subsec:tc-limits}.  The scaling arguments are written in a one-scale notation, $F_1(\nu) \equiv \nu^2/(\nu^2+\omega_0^2)$, where $\omega_0$ denotes a representative polar LO scale.  This is an analytic simplification of the three-mode factor used in the numerical work; it preserves the momentum and frequency regimes responsible for the high- and low-density limits.  We use $\eta$ for the dimensionless coupling to avoid confusion with the gap eigenvalue $\lambda$.

\subsection{High density: \texorpdfstring{$E_{\mathrm{F}}, \omega_{\mathrm{pl}} \gg \omega_0$}{EF >> w0}}

In this limit the relevant momentum transfers are of order $k_{\mathrm{F}}$ and the relevant frequencies obey $|\nu| \lesssim \omega_0 \ll E_{\mathrm{F}}$, so the polarization may be replaced by its static limit, $\Pi \to -m^\ast k_{\mathrm{F}}/\pi^2$, and
\begin{equation}
W(q,\ii\nu)=\frac{4\pi}{\epsilon_\infty}\,
\frac{F_1(\nu)}{q^{2}+F_1(\nu)\,q_{\mathrm{TF}}^{2}},
\qquad
q_{\mathrm{TF}}^{2}=\frac{4 m^\ast k_{\mathrm{F}}}{\pi\,\epsilon_\infty}.
\label{eq:W-static-highn}
\end{equation}
Projecting the normal self-energy on the Fermi surface and performing the angular integrals gives
\begin{equation}
\begin{split}
\Sigma_{\mathrm{N}}(\ii\omega)
&= -\ii\,\frac{m^\ast}{2 k_{\mathrm{F}}\epsilon_\infty}\,
\frac{1}{\beta}\sum_{\omega'}\mathrm{sgn}(\omega')\,F_1(\omega-\omega')\,
\\
&\quad \times
\ln\!\left[1+\frac{4k_{\mathrm{F}}^{2}}{q_{\mathrm{TF}}^{2}\,F_1(\omega-\omega')}\right]\!,
\end{split}
\end{equation}
which at $T=0$ behaves as $\Sigma_{\mathrm{N}} \sim \omega^3$ at low frequency: the static screening removes the phonon contribution to the low-frequency mass renormalization in this regime, consistent with the decrease of $Z$ toward high density in Fig.~\ref{fig:hierarchy}(B).  The Fermi-surface-projected linearized gap equation has the kernel
\begin{equation}
\begin{split}
\Sigma_{\mathrm{A}}(\ii\omega)
&=-\,\eta\,
\frac{\pi}{\beta}\sum_{\omega'}\frac{1}{|\omega'|}\,
F_1(\omega-\omega')\,
\\
&\quad \times
\ln\!\left[1+\frac{4k_{\mathrm{F}}^{2}}{q_{\mathrm{TF}}^{2}\,F_1(\omega-\omega')}\right]
\Sigma_{\mathrm{A}}(\ii\omega'),
\\
\eta&=\frac{m^\ast}{2\pi k_{\mathrm{F}}\epsilon_\infty},
\end{split}
\label{eq:FS-gap-highn}
\end{equation}
so the basic coupling scale decreases as $1/k_{\mathrm{F}}$ (up to the logarithm): this is the high-density flank of the dome.  A two-window Bogoliubov-Tolmachev ansatz~\cite{Bogoliubov59}, $\Sigma_{\mathrm{A}} = \Phi_{\mathrm{low}}$ for $|\omega| < \omega_0$ and $\Phi_{\mathrm{high}}$ for $\omega_0 < |\omega| < E_{\mathrm{F}}$, yields
\begin{equation}
\Phi_{\mathrm{low}} = -\eta\ln 2\,\Phi_{\mathrm{low}}
+\frac{\eta^{2}\ln\frac{E_{\mathrm{F}}}{\omega_0}\,\ln\frac{\omega_0}{T}}
{1+\eta\ln\frac{E_{\mathrm{F}}}{\omega_0}}\,\Phi_{\mathrm{low}} ,
\end{equation}
with the two regimes
\begin{equation}
T_{\mathrm{c}}\;\approx\;
\begin{cases}
\;\omega_0\, \ee^{-1/\eta}, & \eta\ln(E_{\mathrm{F}}/\omega_0)\gtrsim 1,\\[4pt]
\;\omega_0\, \exp\!\left[-\dfrac{1}{\eta^{2}\ln(E_{\mathrm{F}}/\omega_0)}\right], & \eta\ln(E_{\mathrm{F}}/\omega_0)\ll 1 .
\end{cases}
\label{eq:Tc-highn}
\end{equation}
For the STO parameters $\eta \approx 0.154$ at $n = 10^{22}\,\mathrm{cm}^{-3}$ and decreases as $n^{-1/3}$, which makes the high-density collapse of Fig.~\ref{fig:tc} exponentially steep.

\subsection{Low density: \texorpdfstring{$E_{\mathrm{F}} < \omega_0$}{EF < w0}, Fermi-surface projection}

If one insists on projecting onto the Fermi surface in the dilute regime, the frequency transfers are bounded by $E_{\mathrm{F}} \ll \omega_0$ and the polar factor is uniformly small, $F_1 \approx \nu^2/\omega_0^2 \lesssim E_{\mathrm{F}}^2/\omega_0^2$.  The projected gap equation becomes
\begin{equation}
\begin{split}
\Sigma_{\mathrm{A}}(\ii\omega)
&=-\,\eta\,\frac{\pi}{\beta}\sum_{|\omega'|<E_{\mathrm{F}}}
\frac{1}{|\omega'|}\,\frac{(\omega-\omega')^{2}}{\omega_0^{2}}\,
\\
&\quad \times
\ln\!\left[1+\frac{4k_{\mathrm{F}}^{2}}{q_{\mathrm{TF}}^{2}\,F_1(\omega-\omega')}\right]\,
\Sigma_{\mathrm{A}}(\ii\omega'),
\end{split}
\end{equation}
with effective coupling $\bar\eta = \eta\,E_{\mathrm{F}}^{2}/2\omega_0^{2} \propto n$.  Optimizing a two-window ansatz over the dividing frequency $\omega_c$ gives $\omega_c^2/E_{\mathrm{F}}^2 = \bar\eta$ and
\begin{equation}
T_{\mathrm{c}}^{\mathrm{FS}} = E_{\mathrm{F}}\, \ee^{-1}\sqrt{\bar\eta}\;
\exp\!\left[-\frac{1}{\bar\eta^{2}}\right],
\qquad \bar\eta \propto n,
\label{eq:Tc-FS-lown}
\end{equation}
which vanishes faster than any power of $n$: a Fermi-surface theory predicts no observable dilute superconductivity.

\subsection{Low density: full momentum-frequency equation}

The full equation behaves differently because the pair propagator weights states far from the Fermi surface.  In the $\mu \to 0^+$ limit the internal momentum is set by the Matsubara frequency, $p_{\mathrm{typ}}\sim \sqrt{2m^\ast |\omega_m|}$, not by $k_{\mathrm{F}}$, and the long-range kernel produces
\begin{equation}
\int \diff^3p\,\frac{1}{p^2}\,\frac{1}{\omega_m^2+(p^2/2m^\ast)^2}
\;\propto\; \frac{1}{|\omega_m|^{3/2}},
\label{eq:dilute-kernel}
\end{equation}
the dilute kernel of Refs.~\cite{Gastiasoro19-lowdensity,PhanChubukov22-lowdensity}.  The Matsubara sum $T\sum_m |\omega_m|^{-3/2}\sim T^{-1/2}$ converges in the ultraviolet, so the pairing eigenvalue approaches a density-independent function of temperature as $n \to 0$, and the onset survives at $\mu = 0$ for the weak residual low-frequency repulsion appropriate to STO.  In the corresponding one-scale extended Bardeen-Pines model the threshold is $f^\ast \approx 4.9$ and $T_{\mathrm{c}} \propto \omega_0 (f^\ast - f)^2$ near threshold~\cite{PhanChubukov22-lowdensity}; the STO dielectric ratio maps to a value close to the paired side of that problem.  This is consistent with the weakly density-dependent dilute onset computed in Fig.~\ref{fig:tc} and with the collapse of that onset once the kernel is projected or truncated (Fig.~\ref{fig:projection}).

\ZHC{The physical transition at small fixed density is bounded separately by the phase stiffness: even when a non-zero two-particle pairing scale survives as $n \to 0$, the superfluid stiffness vanishes with the carrier density, and the coherent transition cannot exceed}
\begin{equation}
{T_\theta \sim E_{\mathrm{F}}^\ast \approx \frac{E_{\mathrm{F}}}{Z(k_{\mathrm{F}},\ii0)} \sim \frac{n^{2/3}}{Z(k_{\mathrm{F}},\ii0)},}
\label{eq:stiffness}
\end{equation}
or the Bose condensation scale of preformed pairs~\cite{EmeryKivelson95}.  This bound is the gray line of Fig.~\ref{fig:tc}.

\section{Density- and momentum-dependence of soft TO phonon}\label{app:fauque-softmode}

\begin{figure}[!htb]
\centering
\includegraphics[width=0.49\textwidth,clip]{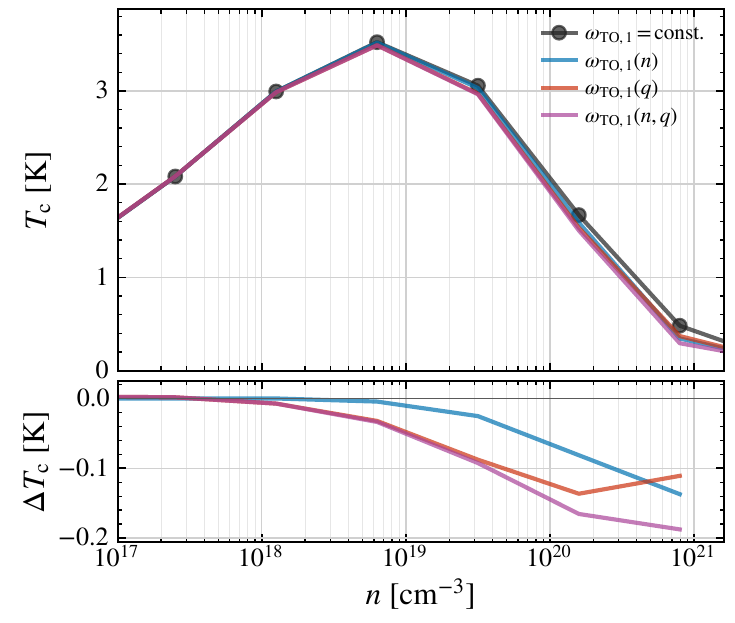}
\caption{\ZHC{Sensitivity of the sc$GW$ pairing-onset dome to density and momentum dependences of the lowest TO phonon.  The upper panel compares the original three-mode result with the three Fauqu\'e-motivated variants (depending on $n$ or $q$ or both); the lower panel shows the change relative to the original sc$GW$ curve.}
\label{fig:fauque-softmode-tc}}
\end{figure}

\ZHC{The calculations in the main text used the density-independent three-pole dielectric factor of Eq.~\eqref{eq:Vtot}.  Fig.~\ref{fig:fauque-softmode-tc} repeats the sc$GW$ $T_{\mathrm{c}}$ scan after adding density and momentum dependence to the lowest TO mode, using the soft-mode data reported by Fauqu\'e \etal~\cite{Fauque25-dipolar-length}, who list doped sample points from $2.4\times10^{18}$ to $1.61\times10^{21}\,\mathrm{cm}^{-3}$, the zone-center soft-mode energies $\omega_{\mathrm{TO}}=1.98$-$17$~meV, and TO velocities $v_{\mathrm{TO}}=8.3$-$5.3$~km/s.  These values are log-interpolated in density inside the table range and held fixed outside it.}

\ZHC{We consider three models. For $\omega_{\mathrm{TO},1}(n)$ we use a relative experimentally determined hardening factor,}
\begin{equation}
h(n)=\frac{\omega_{\mathrm{TO}}^{\mathrm{expt}}(n)}
{\omega_{\mathrm{TO}}^{\mathrm{expt}}(n_{\min})},
\qquad
\omega_{\mathrm{TO},1}(n)=h(n)\,\omega_{\mathrm{TO},1}^{0},
\end{equation}
\ZHC{where $\omega_{\mathrm{TO},1}^{0}=1.24$~meV is the lowest TO pole of the main three-mode model.  The higher TO poles and all LO zeros are kept fixed, and the static dielectric constant is rebuilt from the generalized LST relation, equivalently $\epsilon_0(n)=\epsilon_0^{0}/h^2(n)$.  For $\omega_{\mathrm{TO},1}(q)$ we keep the original zone-center pole and import only the lowest-density dipolar length from the data of Fauqu\'e {\em et al.}}
\begin{equation}
\ell_0=\frac{\hbar v_{\mathrm{TO}}^{\mathrm{expt}}}
{\omega_{\mathrm{TO}}^{\mathrm{expt}}},
\end{equation}
\ZHC{so that the small-$q$ dispersion obeys $\omega_{\mathrm{TO},1}^2(q)\approx(\omega_{\mathrm{TO},1}^{0})^2+(\omega_{\mathrm{TO},1}^{0}\ell_0 q)^2$.  In the numerical calculation this parabola is continued with a rational cap at $0.99\,\omega_{\mathrm{LO},1}$, preserving the small-$q$ slope while keeping the lowest TO pole below the first LO zero.  The $\omega_{\mathrm{TO},1}(n,q)$ curve combines the same relative $q=0$ hardening with the density-dependent $\ell_0(n)=\hbar v_{\mathrm{TO}}^{\mathrm{expt}}(n)/\omega_{\mathrm{TO}}^{\mathrm{expt}}(n)$ and the same $0.99\,\omega_{\mathrm{LO},1}$ cap. 
As an illustration, at $n=1.58\times10^{20}\,\mathrm{cm}^{-3}$, the model
  gives a relative density hardening
 from \(\omega_{\mathrm{TO},1}^{0}=1.24\) to
\(3.67\) meV, and the finite-\(q\) dispersion further gives
\(\omega_{\mathrm{TO},1}(2k_{\mathrm{F}})=9.00\) meV, a factor \(2.45\)
hardening relative to the \(q=0\) value. 
}

\ZHC{Within this construction the effect of the soft-mode hardening is modest.  The three models leave the low-density side and the position of the computed maximum nearly unchanged.  The largest changes occur in the high-density regime: the $n$-only and $q$-only curves lower the sc$GW$ onset by at most about $0.14$~K, while the combined $n,q$ model lowers it by at most about $0.19$~K.  Thus the experimental density and momentum trends of the soft TO pole slightly narrow and lower the overdoped side of the  dome, but they do not remove the order-of-magnitude difference between the sc$GW$ result and the measured transition scale discussed in Sec.~\ref{subsec:tc-limits}.}

\clearpage

\section{Numerical convergence}\label{app:conv}

\begin{figure*}[!htb]
\setlength{\hsize}{\textwidth}
\setlength{\linewidth}{\textwidth}
\includegraphics[width=0.98\textwidth,  clip]{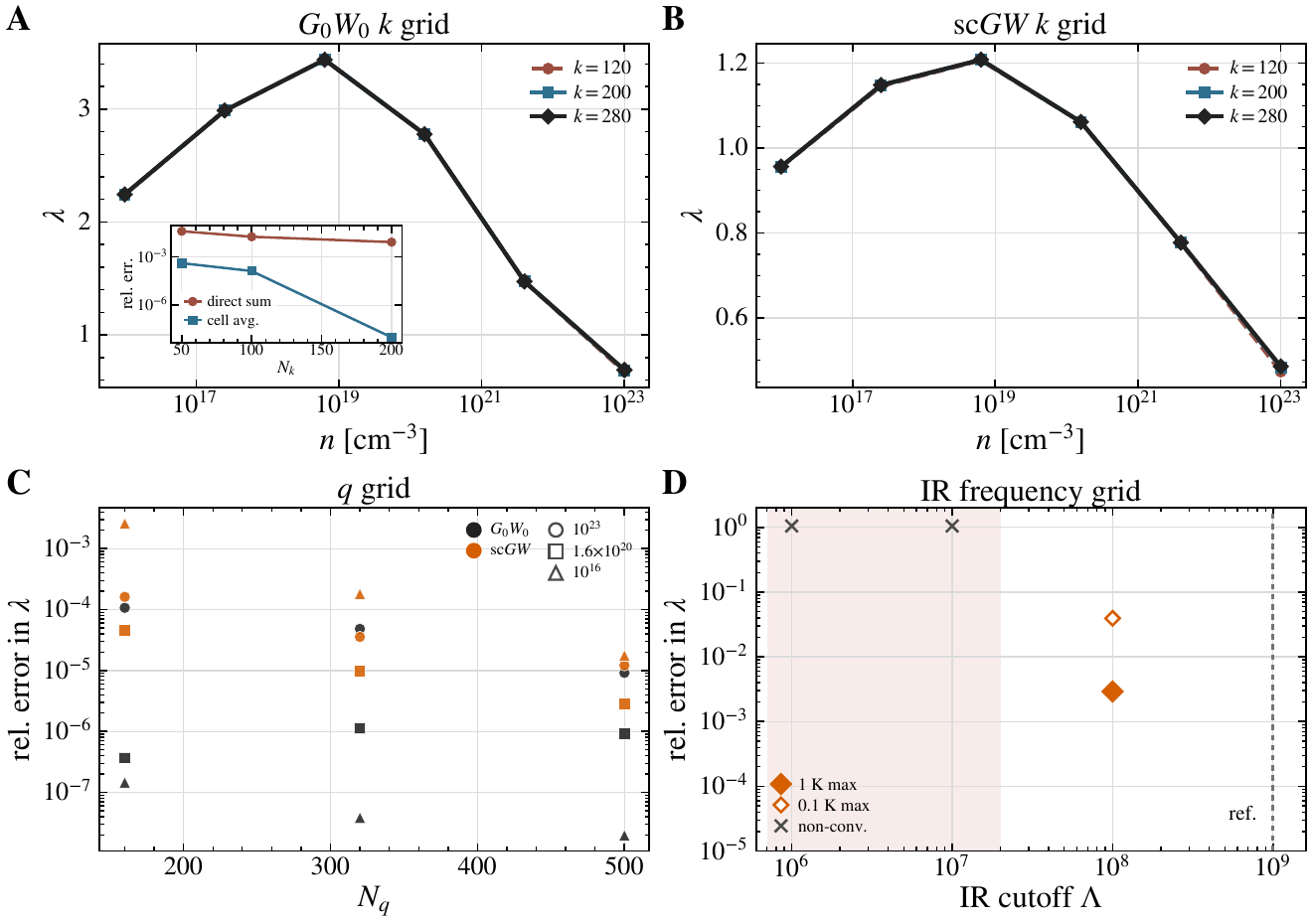}
\caption{Convergence of the full $GW$ pairing eigenvalue.  (A,B) Radial $k$-grid convergence at $T=1$~K for (A) $G_0W_0$ and (B) sc$GW$; the inset in (A) shows the diagonal-block diagnostic, where cell averaging removes the percent-level direct-summation error.  (C) $q$-grid convergence for $G_0W_0$ and sc$GW$ relative to the dense $N_q=750$ reference.  (D) Sparse-IR frequency convergence for sc$GW$ relative to the $\Lambda=10^9$ reference.  The filled diamond is the maximum relative error over the representative $T=1$~K density set and the open diamond is the error at $0.1$~K scan. Crosses with the shaded band mark low-cutoff runs for which either the SCF cycle or the gap equation did not converge.
\label{fig:conv-GW}}
\end{figure*}

Fig.~\ref{fig:conv-GW} documents the grid and basis convergence of the full radial $GW$ solver.  The production radial $k$ settings agree with the denser $k$ grid to better than $0.4\%$ for both $G_0W_0$ and sc$GW$ over the representative density grid, while the production $N_q=500$ grid agrees with the $N_q=750$ reference within $1.8\times10^{-5}$ in the maximum relative error of $\lambda$.  The inset of Fig.~\ref{fig:conv-GW}(A) isolates the remaining singular self-block issue in the diagonal treatment: direct point sampling leaves a percent-level error, whereas the radial-cell average of Appendix~\ref{app:implementation} reduces it to the $10^{-4}$ level on the same grids.

The sparse-IR cutoff is the most delicate convergence variable in the full sc$GW$ calculation: the small cutoffs $\Lambda=10^6$ and $10^7$ often fail either the SCF cycle or the Davidson gap solve [Fig.~\ref{fig:conv-GW}(D)].  With $\Lambda=10^8$, the $T=1$~K sc$GW$ eigenvalue differs from the $\Lambda=10^9$ reference by at most $2.9\times10^{-3}$ over the representative density set.  At $T=0.1$~K the same comparison is more stringent: the maximum sc$GW$ difference is $3.9\%$, while the corresponding $G_0W_0$ maximum is $0.46\%$.  We therefore use $\Lambda=10^9$ for the production data and the low-temperature scans.
\paragraph*{One-mode compression check.}
The main text uses the explicit three-pole lattice dielectric factor.  As a compression check, we repeated the same figure workflow with the single LST-consistent doublet defined by the log-strength average of $S(\ii\nu)=1-F_3(\nu)$, $\omega_{\mathrm{LO,eff}}=84.26$~meV and $\omega_{\mathrm{TO,eff}}=1.38$~meV.
Other simple moment fitting choices give LO scales between $75.67$ and $90.26$~meV, so this is a defined compression convention rather than a distinct physical branch.  At $T=1$~K the $G_0W_0$ and $GW_0$ density scans differ from the three-mode curves only at the percent level over the production grid; the sc$GW$ low-density eigenvalue is more sensitive because the $T=1$~K point sits near the pairing threshold.  The full $T_{\mathrm{c}}$ scan remains qualitatively unchanged: the sc$GW$ peak is $3.52$~K in the three-mode model and $3.93$~K in the compressed one-mode model.  Thus the conclusions of the main text do not rely on the one-mode compression, and the production figures use the more faithful three-mode dielectric factor.

\FloatBarrier
\bibliography{refs}

\end{document}